\documentclass[reprint, double column, superscriptaddress,prl, showkeys]{revtex4-1}

\usepackage{float}
\usepackage{graphicx,epsfig}
\usepackage{amssymb}
\usepackage{amsmath}
\usepackage{bm}
\usepackage{braket}

\usepackage{makecell}
\usepackage{textcomp}
\usepackage{color}
\usepackage{pgffor}
\usepackage{pdfpages}

\usepackage{soul}

\usepackage[bookmarksnumbered,bookmarksopen]{hyperref}
\hypersetup{
   colorlinks=true,       
}

\usepackage{BOONDOX-cal}

\makeatletter

\AtBeginDocument{\let\LS@rot\@undefined}
\makeatother
\newif\ifarXiv
\arXivtrue

\usepackage{soul}

\begin{document}
\setcounter{page}{1}

\title[]{Even-denominator fractional quantum Hall states with spontaneously broken rotational symmetry}
\author{Chengyu \surname{Wang}}
\thanks{E-mail: chengyuw@princeton.edu and shayegan@princeton.edu}
\author{A. \surname{Gupta}}
\author{S. K. \surname{Singh}}
\author{C. T. \surname{Tai}}
\author{L. N. \surname{Pfeiffer}}
\author{K. W. \surname{Baldwin}}
\affiliation{Department of Electrical and Computer Engineering, Princeton University, Princeton, New Jersey 08544, USA}
\author{R. \surname{Winkler}}
\affiliation{Department of Physics, Northern Illinois University, DeKalb, Illinois 60115, USA}
\author{M. \surname{Shayegan}}
\thanks{E-mail: chengyuw@princeton.edu and shayegan@princeton.edu}
\affiliation{Department of Electrical and Computer Engineering, Princeton University, Princeton, New Jersey 08544, USA}

\date{\today}

\begin{abstract}
The interplay between the fractional quantum Hall effect and nematicity is intriguing as it links emerging topological order and spontaneous symmetry breaking. Anisotropic fractional quantum Hall states (FQHSs) have indeed been reported in GaAs quantum wells but only in tilted magnetic fields, where the in-plane field explicitly breaks the rotational symmetry. Here we report the observation of FQHSs with highly anisotropic longitudinal resistances in purely perpendicular magnetic fields at even-denominator Landau level fillings $\nu=5/2$ and 7/2 in ultrahigh-quality GaAs two-dimensional hole systems. The coexistence of FQHSs and spontaneous symmetry breaking at half fillings signals the emergence of nematic FQHSs which also likely harbor non-Abelian quasiparticle excitations. By gate tuning the hole density, we observe a phase transition from an anisotropic, developing FQHS to an isotropic composite fermion Fermi sea at $\nu=7/2$. Our calculations suggest that the mixed orbital components in the partially occupied Landau level play a key role in the competition and interplay between topological and nematic orders.
\end{abstract}

\maketitle

\section{Introduction}
\begin{figure}[b!]
  \begin{center}
    \psfig{file=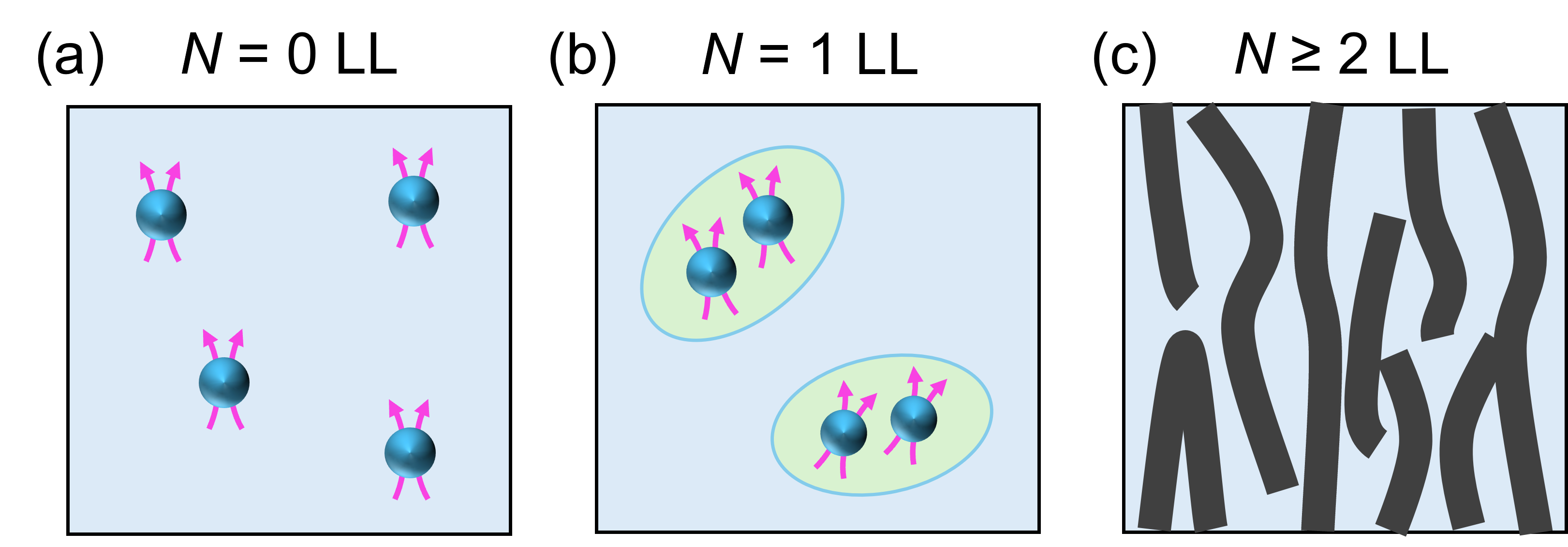, width=0.48 \textwidth}
  \end{center}
  \caption{\label{schematic} 
   Ground states at half-filled LLs: (a) CF Fermi sea ($N=0$), (b) paired FQHS ($N=1$), and (c) nematic phase ($N\geq2$).
   }
  \label{fig:schematic}
\end{figure}

Fractional quantum Hall states (FQHSs) are fascinating examples of interaction phenomena and topological order in condensed matter physics. Most FQHSs are observed at odd-denominator Landau level (LL) filling factors, following the Jain sequence \cite{Tsui.PRL.1982, Jain.PRL.1989, Jain.Book.2007}. These states occur in the ground-state ($N=0$) LLs where short-range Coulomb interaction dominates, and can be explained as integer quantum Hall states of weakly interacting composite fermions (CFs). The ground state at a half-filled $N=0$ LL (e.g., at $\nu=1/2$) is a compressible Fermi sea of CFs; see Fig. \ref{fig:schematic}(a). In excited ($N\geq1$) LLs, the nodal structure of the wavefunction softens the short-range Coulomb interaction, leading to a variety of exotic states that are not favored in the $N=0$ LLs \cite{Willett.PRL.1987, Lilly.PRL.1999, Du.SSC.1999, Eisenstein.PRL.2002, Kumar.PRL.2010, Fu.PRL.2020}. These include FQHSs at even-denominator filling factors, first observed at a half-filled $N=1$ LL ($\nu=5/2$) of GaAs two-dimensional electron systems (2DESs) \cite{Willett.PRL.1987}. At $\nu=5/2$, the CF Fermi sea is no longer stable as the CFs undergo a Bardeen–Cooper–Schrieffer-like pairing instability and condense into a FQHS [Fig. \ref{fig:schematic}(b)]. The prime candidates for the $\nu=5/2$ FQHS are believed to host non-Abelian anyons, and be potentially useful for fault-tolerant topological quantum computation \cite{Moore.NPB.1991, Nayak.RMP.2008, Banerjee.Nature.2018, Halperin.Book.2020, Willett.PRX.2023}. Even-denominator FQHSs at half fillings of the $N=1$ LLs were also reported in other material platforms, including semiconductor heterostructures \cite{Falson.NatPhys.2015, Hossain.PRL.2018} and atomically-thin 2D materials \cite{Ki.NanoLett.2014, Zibrov.Nature.2017, Li.Science.2017, Huang.PRX.2022, Shi.NatNano.2020}. 

In $N\geq2$ LLs, stripe or electronic versions of liquid-crystal-like phases with spontaneously broken rotational symmetry emerge near half fillings \cite{Lilly.PRL.1999, Du.SSC.1999, Fu.PRL.2020, Koulakov.PRL.1996, Fogler.PRB.1996, Moessner.PRB.1996, Fradkin.PRB.1999, Fradkin.PRL.2000, Fradkin.ARCMP.2010}. These states exhibit anisotropic in-plane transport coefficients with no quantized Hall plateau, and were initially predicted by Hartree-Fock theories to stem from unidirectional charge-density waves consisting of stripes with alternating integer $\nu$ \cite{Koulakov.PRL.1996, Fogler.PRB.1996, Moessner.PRB.1996, Sammon.PRB.2019}. At finite temperatures, and in the presence of quantum fluctuations and disorder, the stripe order can be disrupted, leading to nematic phases [Fig. \ref{fig:schematic}(c)] \cite{Fradkin.PRB.1999, Fradkin.PRL.2000, Fradkin.ARCMP.2010}. Alternatively, in a more general picture, nematic orders can also arise from Pomeranchuk instability of Fermi seas \cite{Oganesyan.PRB.2001, Lee.PRL.2018}.

\begin{figure*}[t]
  \begin{center}
    \psfig{file=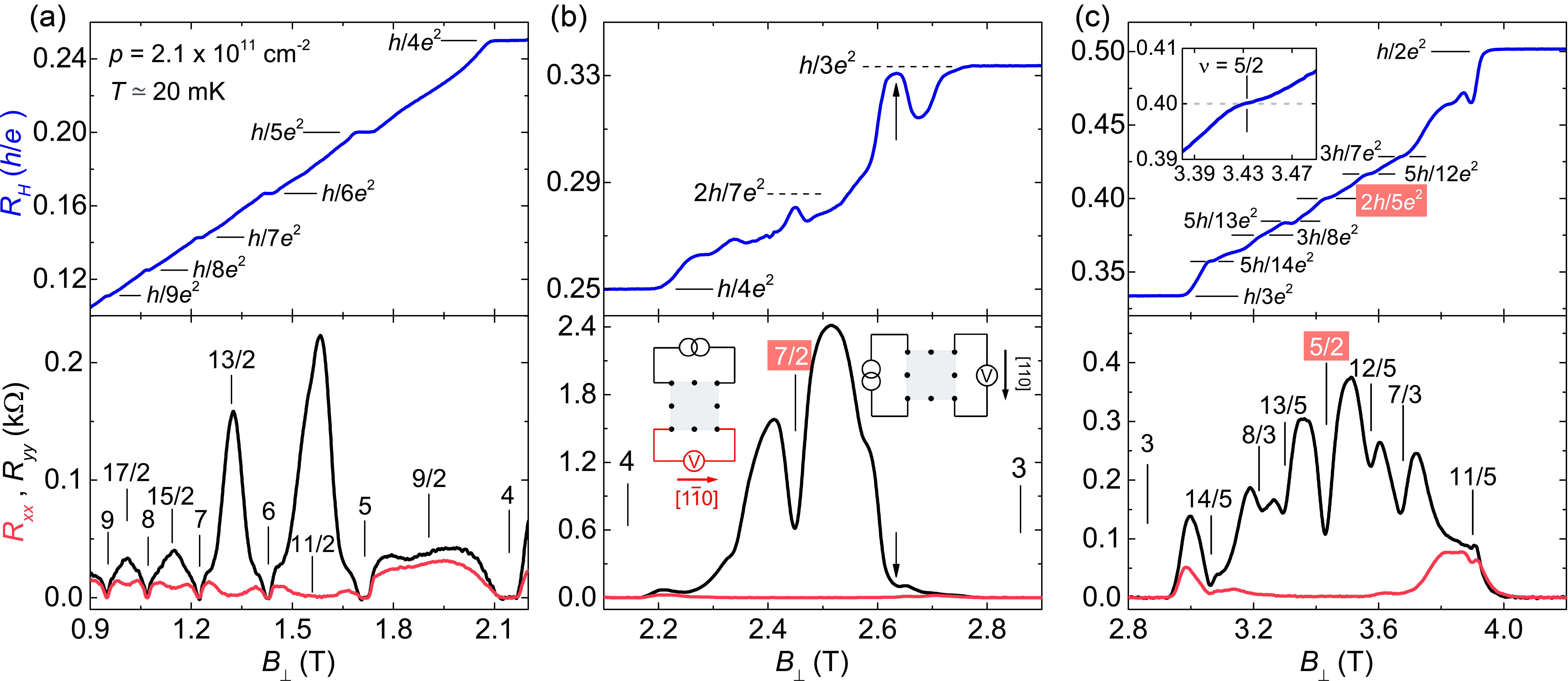, width=1\textwidth}
  \end{center}
  \caption{\label{base} 
   Hall resistance $R_H$ (top panels) and longitudinal resistances $R_{xx}$ and $R_{yy}$ (bottom panels) for our ultrahigh-quality 2DHS; sample A. The sample has a quantum well width of 20 nm and a 2D hole density of 2.1$\times10^{11}$ cm$^{-2}$. Numerous anisotropic phases are observed near half fillings. (a) Anisotropic phases, without Hall plateaus, in high LLs near $\nu=11/2$, 13/2, 15/2, and 17/2. Note that the 2DHS is isotropic at $\nu=9/2$. (b) A developing FQHS with large transport anisotropy at even-denominator $\nu=7/2$. Insets in the lower panel show the circuit configurations for $R_{xx}$ and $R_{yy}$ measurements. (c) Developing FQHSs with large transport anisotropy at even-denominator $\nu=5/2$ and at numerous odd-denominator fillings 8/3, 13/5, 12/5, and 7/3.
   }
  \label{fig:base}
\end{figure*}

While the vast majority of FQHSs are isotropic, FQHSs with significant longitudinal transport anisotropy were observed in the $N=1$ LL of GaAs 2DESs at $\nu=7/3$ and 5/2 in tilted magnetic fields \cite{Xia.NatPhys.2011, Liu.PRB.2013}. Another example of an anisotropic FQHS was reported in the half-filled $N=1$ LL of an AlAs 2DES with large band-mass anisotropy and an applied uniaxial strain \cite{Hossain.PRL.2018}. These experiments suggest that FQHSs may exhibit nematic order, forming the so-called nematic FQHSs. These states are very unusual, because unlike the conventional nematic states which are understood in terms of Landau’s symmetry-breaking theory and described by traditional order parameter, nematic FQHSs possess both topological and nematic order \cite{Wen.AP.1995}. Nematicity in FQHSs has indeed attracted significant interest because of its connection to the geometrical degree of freedom in FQHSs and the intriguing coexistence of topological order with broken symmetry \cite{Haldane.PRL.2011, Mulligan.PRB.2011, Maciejko.PRB.2013, You.PRX.2014, Wan.PRB.2016, Yang.PRL.2017, Regnault.PRB.2017, Santos.PRX.2019, Yang.PRR.2020, Ye.PRB.2024, Pu.PRL.2024}. Nematic FQHSs are theoretically predicted to exhibit a finite charge gap and a vanishing neutral gap in the long-wavelength limit \cite{Maciejko.PRB.2013, You.PRX.2014}. Anisotropic FQHSs reported in the $N=1$ LL are likely candidates for nematic FQHSs \cite{Xia.NatPhys.2011, Liu.PRB.2013, Hossain.PRL.2018, Wang.PRL.2023}; however, their observation has until now required the application of external symmetry-breaking fields, such as in-plane magnetic fields or uniaxial strain, which \textit{explicitly} break the rotational symmetry. In a purely perpendicular magnetic field, an unusual phase transition from a nearly isotropic FQHS to a \textit{compressible} nematic phase, induced by hydrostatic pressure, was reported, indicating the proximity of nematic order to FQHSs \cite{Samkharadze.NatPhys.2016}. However, experimental evidence for a nematic FQHS with \textit{spontaneously} broken rotational symmetry has been lacking until now.

Here we report the observation of anisotropic FQHSs at \textit{even-denominator} fillings $\nu=5/2$ and 7/2 in ultrahigh-quality GaAs 2D \textit{hole} systems (2DHSs). Remarkably, these states are observed in the absence of an external symmetry-breaking field, suggesting the spontaneous rotational symmetry breaking of candidate non-Abelian FQHSs. The emergence of anisotropic FQHSs at $\nu=5/2$ and 7/2 are evinced by pronounced minima superimposed on a highly anisotropic longitudinal resistance background, accompanied by developing quantized Hall plateaus. These anisotropic FQHSs exhibit a unique combination of characteristics of both half-filled $N=1$ LL and $N\geq2$ LLs, revealing the coexistence of CF paring and nematic order; see Figs. \ref{fig:schematic}(b,c). We also observe qualitatively similar, albeit weaker, signatures of anisotropic FQHSs at odd-denominator fillings flanking $\nu=5/2$. At higher LL fillings ($\nu>4$), we observe numerous anisotropic phases without any FQHS features, consistent with what one would expect for high ($N\geq2$) LLs.

\section{Methods}
We studied ultrahigh-quality 2DHSs confined to GaAs quantum wells grown on GaAs (001) substrates by molecular beam epitaxy \cite{Chung.Natmater.2021, Chung.PRM.2022, Gupta.PRM.2024}. Our magneto-transport measurements were performed on $4\times 4$ mm$^2$ van der Pauw geometry samples, with alloyed In:Zn contacts at the four corners and side midpoints. We studied seven samples from six different wafers; see Table I in Supplemental Material (SM) \cite{SM} for sample parameters. Sample C was fitted with a Ti/Au front gate and In back gate, allowing for \textit{in-situ} tuning of the 2D hole density. The samples were cooled in two dilution refrigerators with base temperatures of about 20 mK and 30 mK. We measured the longitudinal in-plane resistances ($R_{xx}$ and $R_{yy}$) along two perpendicular directions and the Hall resistance ($R_H$) using the low-frequency, lock-in amplifier techniques.

It is worth noting that Hall measurements for anisotropic phases are challenging because of the non-uniform current distribution and significant mixing of $R_{xx}$ signal into the Hall meausrements \cite{Xia.NatPhys.2011, Hossain.PRL.2018, Fu.PRL.2020}. One effective method to minimize the effect of $R_{xx}$ mixing is by averaging $|R_{xy}|$ taken for opposite polarities of $B_{\perp}$. Because our magnet only has the capability of applying large $B_{\perp}$ in one direction, we used an alternative method to minimize the effect of $R_{xx}$ mixing \cite{Goldman.PRL.1988}: we measured Hall resistances with two circuit configurations (rotated by 90$^{\circ}$) and took their average value as $R_H$.

\section{Results}
Figure \ref{fig:base} presents Hall and longitudinal resistance data measured as a function of $B_{\perp}$. Longitudinal resistances $R_{xx}$ and $R_{yy}$ were measured along the [1$\bar{1}$0] and [110] crystallographic directions, respectively (see insets in Fig. \ref{fig:base}(b) lower part for circuit geometries). Panel (a) shows data for $\nu\geq4$. We observe anisotropic phases at half fillings $\nu=11/2$, 13/2, 15/2, and 17/2, evinced by broad dips in $R_{xx}$ and peaks in $R_{yy}$. These anisotropic states resemble the stripe/nematic phases reported in GaAs 2DESs at high $N\geq2$ LLs \cite{Lilly.PRL.1999, Du.SSC.1999}. Near $\nu=9/2$, the 2DHS exhibits a nearly isotropic behavior. As we detail in the Discussion later, the isotropic phase at $\nu=9/2$ is well explained by our LL analysis. Our data in Fig. \ref{fig:base}(a) are consistent with previous reports on high-quality GaAs 2DHSs with similar parameters \cite{Manfra.PRL.2007}. We note that, for $\nu>4$, the Hall resistance $R_H$ remains smooth and linear as a function of $B_{\perp}$ between the quantized Hall plateaus at integer fillings.

\begin{figure*}[t]
  \begin{center}
    \psfig{file=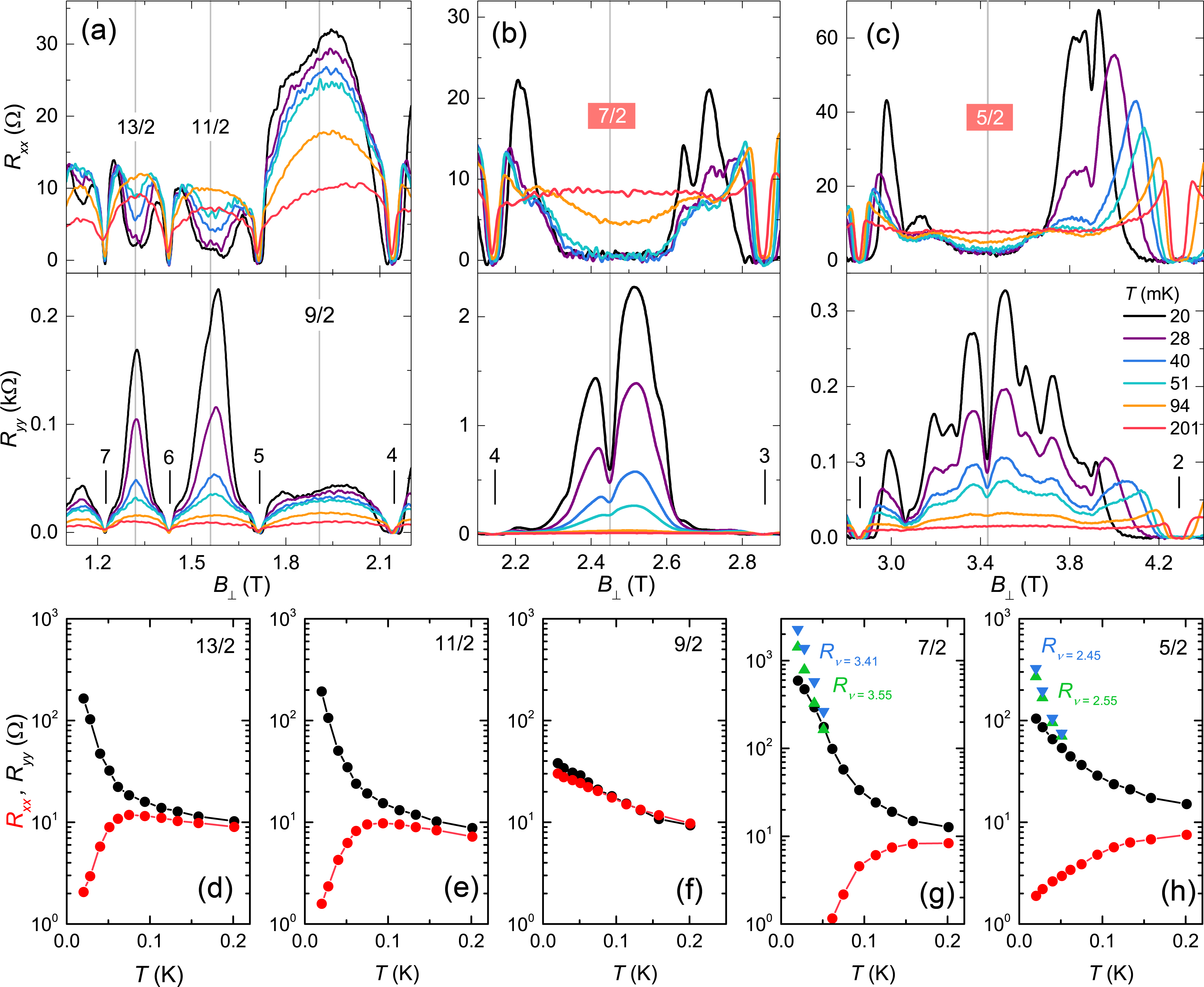, width=1\textwidth}
  \end{center}
  \caption{\label{fig:Tdep}
    Temperature dependence data; sample A. (a-c) $R_{xx}$ and $R_{yy}$ vs $B_{\perp}$ traces taken at different temperatures for: (a) $\nu\geq4$, (b) $4\geq\nu\geq3$, and (c) $3\geq\nu\geq2$. (d-h) Temperature dependence of $R_{xx}$ and $R_{yy}$ at half fillings: (d) $\nu=13/2$, (e) 11/2, (f) 9/2, (g) 7/2, and (h) 5/2. Blue and green triangles show $R_{xx}$ values on the flanks of $\nu=5/2$ and 7/2 at the indicated filling factors.
    }
  \label{fig:Tdep}
\end{figure*}

Near $\nu=7/2$, the 2DHS exhibits strong anisotropy, with $R_{xx}\simeq0$ and $R_{yy}\simeq2$ k$\Omega$ [Fig. \ref{fig:base}(b)]. Unlike the anisotropic phases observed at higher $\nu$, a sharp minimum in $R_{yy}$ is observed at $\nu=7/2$, superimposed on a resistive background. While $R_H$ exhibits a complex behavior because of the non-uniform current distribution within the highly anisotropic phase, its value approaches $2h/7e^2$ to within $2\%$ at $\nu=7/2$. These observations suggest a developing, anisotropic FQHS at $\nu=7/2$. At $B_{\perp}\simeq 2.64$ T ($\nu\simeq3.3$), $R_{H}$ shows a reentrant behavior towards the $h/3e^2$ plateau, indicating the emergence of a developing reentrant integer quantum Hall state. This is also supported by $R_{xx}$ and $R_{yy}$ data where developing minima are seen at the same $B_{\perp}$ position; these are more clearly resolved in measurements employing different contact configurations; see SM Fig. S1 \cite{SM}. Such reentrant integer quantum Hall behavior may signal the Wigner crystallization of multi-electron bubbles \cite{Eisenstein.PRL.2002, Koulakov.PRL.1996, Moessner.PRB.1996}.

Data presented in Fig. \ref{fig:base}(c) for $3>\nu>2$ reveal our strongest evidence for anisotropic FQHSs. At $\nu=5/2$, a sharp $R_{yy}$ minimum, accompanied with a developing Hall plateau centered at $R_H=2h/5e^2$, signals an even-denominator FQHS. Similar to what we see at $\nu=7/2$, the 5/2 FQHS also exhibits anisotropic transport ($R_{xx}<<R_{yy}$). Qualitatively similar features are seen at odd-denominator fillings $\nu=8/3$, 13/5, 12/5, and 7/3, suggesting developing anisotropic FQHSs at these fillings also. These signatures of anisotropic FQHSs at odd- and even-denominator fillings are reproducible in multiple samples from different wafers; see Fig. \ref{fig:density} and SM Figs. S4 and S5 for more data \cite{SM}. Importantly, \textit{the anisotropic FQHSs in our 2DHS are observed in the absence of any external symmetry-breaking field} (e.g. an in-plane magnetic field). Our results therefore provide strong evidence for spontaneously broken rotational symmetry in these FQHSs, signaling the emergence of \textit{nematic FQHSs}.

We observe that the easy axis direction is along [1$\bar{1}$0] for both the nematic FQHSs at $\nu=$ 5/2 and 7/2 as well as for stripe/nematic phases in higher Landau levels (e.g., $\nu=$ 11/2, 13/2, ...), suggesting a common rotational symmetry breaking mechanism for these states. In GaAs 2DESs, the majority of stripe/nematic phases in $N\geq2$ LLs exhibit an easy axis along the [110] direction, although examples of the easy axis along [1$\bar{1}$0] have been reported under specific conditions \cite{Zhu.PRL.2002, Pollanen.PRB.2015}. The origin of the internal symmetry breaking field that favors a certain crystal direction over the other remains an unresolved question.

Figure \ref{fig:Tdep} captures the temperature dependence of the observed nematic FQHSs and anisotropic phases. Note the differing y-axis scales in Figs. \ref{fig:Tdep}(a-c). The temperature dependence of $R_{xx}$ and $R_{yy}$ at various half fillings is further highlighted in Figs. \ref{fig:Tdep}(d-h). At $\nu=9/2$ [Fig. \ref{fig:Tdep}(f)], we observe nearly isotropic transport with both $R_{xx}$ and $R_{yy}$ gradually increasing with decreasing $T$ down to $T\simeq 20$ mK. In contrast to the isotropic behavior near $\nu=9/2$, at $\nu=13/2$, 11/2, 7/2, and 5/2, $R_{xx}$ and $R_{yy}$ significantly deviate from each other and exhibit opposite trends at low $T$, indicating the emerging nematic order [Figs. \ref{fig:Tdep}(d,e,g,h)]. Furthermore, $R_{yy}$ minima, superimposed on a rising background, rapidly develop at $\nu=5/2$ and 7/2. Linear fits to Arrhenius plots of the relative depth of these $R_{yy}$ minima yield energy gap estimates of approximately 55 mK and 76 mK for the $\nu=5/2$ and 7/2 FQHSs, respectively; see SM Fig. S2 for details \cite{SM}. The simultaneous emergence of FQHS and nematic order strongly supports our interpretation of these states as developing nematic FQHSs. We note that the anisotropic transport near $\nu=7/2$ and 5/2 persists to higher temperatures ($\simeq 150$ mK) when the FQHS features disappear, similar to what was reported for anisotropic FQHSs in tilted $B$ \cite{Xia.NatPhys.2011, Liu.PRB.2013, Wang.PRL.2023}.  

\begin{figure}[t!]
  \begin{center}
    \psfig{file=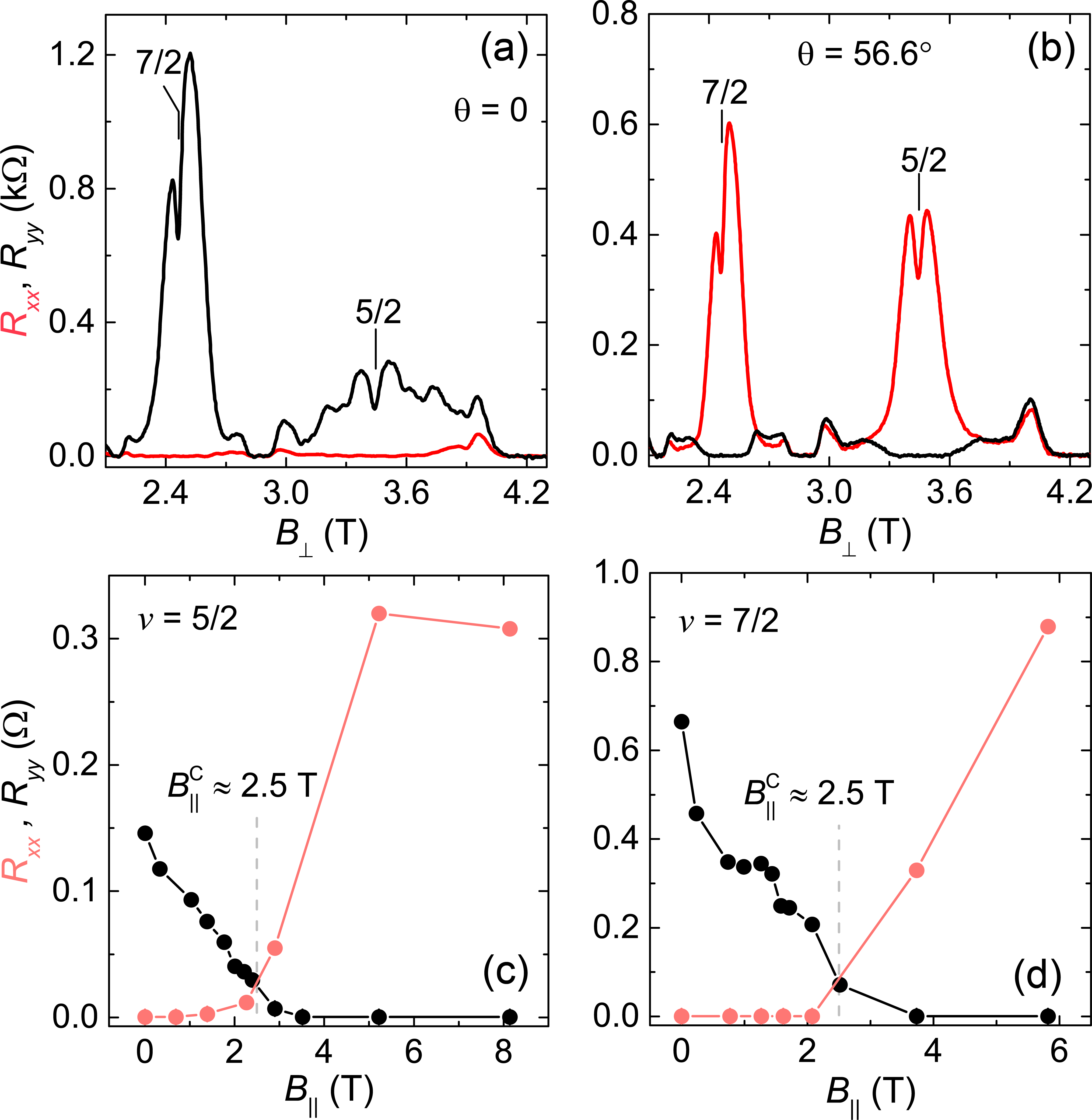, width=0.49\textwidth}
  \end{center}
  \caption{\label{tilt}
   Impact of in-plane magnetic field ($B_{\parallel}$) on the nematic FQHSs; sample B, $T\simeq$ 30 mK. This sample has a quantum well width of 20 nm and 2D hole density of $2.1\times10^{11}$ cm$^{-2}$. (a, b) $R_{xx}$ and $R_{yy}$ vs $B_{\perp}$ traces measured at (a) zero tilt and (b) tilt angle $\theta=56.6^{\circ}$. (c, d) $R_{xx}$ and $R_{yy}$ at $\nu=5/2$ and 7/2 as a function of $B_{\parallel}$, revealing the switching of the hard and easy axes at a critical in-plane field of $B^C_{\parallel}\simeq2.5$ T. Inset in (a): A schematic of the experimental setup. The sample is mounted on a rotating stage to support \textit{in-situ} tilt; $\theta$ is the angle between $B$ and $B_{\perp}$. By tilting the sample, a finite $B_{\parallel}=B\times\sin{\theta}$ is introduced along the [1$\Bar{1}$0] crystallographic direction.}
  \label{fig:tilt}
\end{figure}

Applying an in-plane magnetic field, $B_{\parallel}$, provides further insight into the nematic order at half fillings. In Figs. \ref{fig:tilt}(a,b), we show $R_{xx}$ and $R_{yy}$ data in perpendicular and tilted $B$, measured for sample B at $T\simeq30$ mK; see Fig. S6 in SM for more data at different $\theta$ \cite{SM}. This sample was mounted on a rotating stage to support \textit{in-situ} tilt; see Fig. \ref{fig:tilt}(a) inset. With the sample tilted, $B_{\parallel}$ is applied along [1$\bar{1}$0] (parallel to $R_{xx}$) direction. At $\theta=0$, we observe developing anisotropic FQHSs at $\nu=5/2$ and 7/2 with $R_{xx}<<R_{yy}$. At $\theta=56.6^{\circ}$, developing FQHSs survive, but the hard and easy axes of resistive anisotropy are interchanged, resulting in $R_{xx}>>R_{yy}$. We plot $R_{xx}$ and $R_{yy}$ at $\nu=5/2$ and 7/2 as a function of $B_{\parallel}$ [Figs. \ref{fig:tilt}(c,d)]. A critical in-plane field, $B_{\parallel}^C\simeq$ 2.5 T, is identified where the 2DHS becomes nearly isotropic at both $\nu=5/2$ and 7/2. The orientation of the hard axis at large $B_{\parallel}$ is aligned with the direction of the applied $B_{\parallel}$, consistent with what was observed in GaAs 2DESs \cite{Lilly.PRL.1999.tilt, Pan.PRL.1999.tilt, Liu.PRB.2013, Cooper.PRB.2002}. Such tilt-induced anisotropy is also consistent with Hartree-Fock calculations of the unidirectional charge density wave orientation energy induced by a tilted magnetic field \cite{Jungwirth.PRB.1999}.

\begin{figure*}[t]
  \begin{center}
    \psfig{file=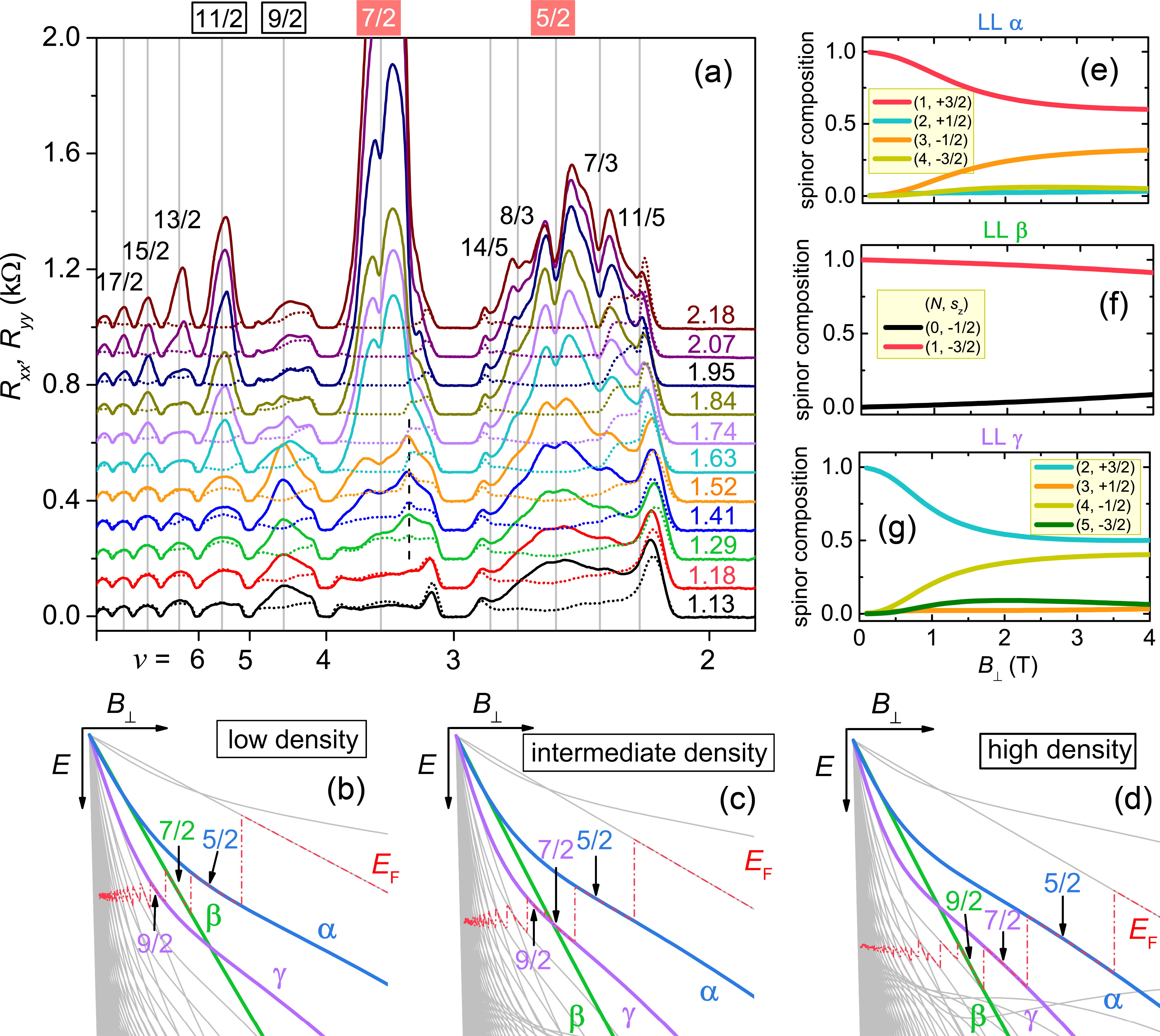, width=0.9 \textwidth}
  \end{center}
  \caption{\label{fig:density}
    Gate-tuned phase transitions at half fillings; sample C, $T\simeq$ 20 mK. This sample has a quantum well width of 20 nm and an as-grown density of 1.74$\times10^{11}$ cm$^{-2}$. (a) $R_{xx}$ (along [1$\bar{1}$0], dotted) and $R_{yy}$ (along [110], solid) traces measured at different densities. Each trace is vertically shifted by 0.1 k$\Omega$ for clarity, with its corresponding 2D hole density ($p$), in units of 10$^{11}$ cm$^{-2}$, indicated on the right. While the longitudinal transport at $\nu=5/2$ is anisotropic in the density range of 1.13 to 2.18, transitions between isotropic and anisotropic phases are observed as a function of density at half fillings in higher LLs, e.g, at $\nu=7/2$, 9/2, and 11/2. (b-d) Calculated energy vs $B_{\perp}$ LL fan diagrams at low ($p=0.74$), intermediate ($p=1.04$), and high ($p=1.74$) densities. Three LLs ($\boldsymbol{\alpha}$, $\boldsymbol{\beta}$, and $\boldsymbol{\gamma}$ levels) of interest are highlighted. Other LLs are shown in light grey. Red dash-dotted line traces the Fermi energy $E_F$. Black arrows mark the positions of $\nu=5/2$, 7/2, and 9/2. (e-g) Spinor compositions of $\boldsymbol{\alpha}$, $\boldsymbol{\beta}$, and $\boldsymbol{\gamma}$ LLs.}
  \label{fig:density}
\end{figure*}

Next, we address the origin of the nematic FQHSs we observe at $\nu=5/2$ and 7/2 in our 2DHSs. FQHSs at half fillings in different 2DESs are predominantly observed in the $N=1$ LLs \cite{Willett.PRL.1987, Falson.NatPhys.2015, Hossain.PRL.2018, Zibrov.Nature.2017, Li.Science.2017, Huang.PRX.2022, Shi.NatNano.2020}. On the other hand, the unique property of the 5/2 and 7/2 FQHSs in our 2DHSs is the spontaneous breaking of rotational symmetry. Such electronic nematic order is reminiscent of the anisotropic phases observed in higher $N\geq2$ LLs in GaAs 2DESs \cite{Lilly.PRL.1999, Du.SSC.1999}. Indeed, the LLs of GaAs 2DHSs are very different compared to electrons. Because of the mixing between the heavy-hole and light-hole states in $B_{\perp}$ and significant spin-orbit coupling, hole LLs have a multi-orbital nature \cite{Winkler.Book.2003}. In addition, hole LLs are highly nonlinear as a function of $B_{\perp}$ and exhibit numerous crossings that can be tuned to coincidence with the Fermi energy, $E_F$ \cite{Liu.PRB.2014, Lupatini.PRL.2020, Ma.PRL.2022, Wang.PRL.2023}. To shed light on the nature of LLs that host the nematic FQHSs, we studied the evolution of different states at half fillings with varying hole densities, and compare our data with the calculated LL fan diagrams for our sample.

Focusing first on the experimental data, in Fig. \ref{fig:density}(a), we present $R_{xx}$ and $R_{yy}$ data for sample C at different densities. We tune the hole density from 1.13 to 1.84 (in units of $10^{11}$ cm$^{-2}$, which we use throughout the remainder of the manuscript) using front and back gates while keeping the charge distribution symmetric. Higher hole densities ($>1.84$) are achieved by further changing the front gate voltage only, and the charge distribution becomes slightly asymmetric. Near $\nu=5/2$, the 2DHS is anisotropic in the density range $p=$ 1.13 to 2.18. The FQHS features at $\nu=5/2$ and 7/3 are absent at $p=1.13$ likely because of the very low 2DHS density, and gradually develop with increasing density. No sharp transition is seen for FQHSs between $\nu=3$ and 2. Near $\nu=7/2$, on the other hand, the 2DHS is isotropic at low densities but becomes highly anisotropic at high densities. A sharp isotropic-to-anisotropic transition occurs at $p\simeq1.3$. Interestingly, an $R_{yy}$ minimum at $\nu=7/2$ appears simultaneously with the onset of anisotropy. We also observe isotropic-to-anisotropic transitions in higher LLs, e.g., at $\nu=11/2$. The behavior near $\nu=9/2$ is rather different: The 2DHS is anisotropic at low densities but becomes isotropic at $p\simeq2.0$. At $p=2.18$, the 2DHS becomes slightly anisotropic near $\nu=9/2$. The anisotropy at $\nu=9/2$, however, is much weaker than that at nearby half fillings, e.g., at $\nu=7/2$ and 11/2, where $R_{xx}$ show deep minima. No sharp $R_{yy}$ minimum is observed at half fillings other than 5/2 and 7/2.

\section{Discussion}
In order to understand the origin of the different ground states we observe at different fillings, and their evolution with density, we present calculated LL fan diagrams for our 2DHS; for more details of the LL calculations and results, see SM \cite{SM}. These LL diagrams [Figs. \ref{fig:density}(b-d)], together with the spinor compositions of the relevant LLs [Figs. \ref{fig:density}(e-g)] shed significant light on the emergence of the different states in the 2DHS. The calculations are performed for our 2DHS (sample C) in a purely $B_{\perp}$ using the multiband envelope function approximation and the $8\times8$ Kane Hamiltonian \cite{Winkler.Book.2003}. We also use axial approximation so that we can decompose each LL into the basis of four spinors with $s_z=\pm3/2$ and $\pm1/2$. The spinor composition is directly related to the orbital composition of each LL \cite{Winkler.Book.2003, Wang.PRL.2023}.  In Figs. \ref{fig:density}(b-d), we present three representative LL fan charts, calculated at different densities $p=0.74$, 1.04, and 1.74. The three most relevant LLs ($\boldsymbol{\alpha}$, $\boldsymbol{\beta}$, and $\boldsymbol{\gamma}$) are highlighted. The spinor compositions for these LLs, shown in Figs. \ref{fig:density}(d-f), are calculated for $p=1.74$. Our calculations show that, for a given quantum well width, density has a negligible effect on these results \cite{SM}. They are therefore representative for the whole range of densities we studied for sample C. We also note that, although there are some discrepancies, our LL calculations qualitatively capture the main features in the $E$ vs $B_{\perp}$ LL fan charts, e.g., the nonlinearity in $E$ vs $B_{\perp}$ and the numerous LL crossings at finite $B_{\perp}$.

Before we elaborate on the origin of different states at different fillings, we summarize the nature of the three LLs which are most relevant to our studies, the $\boldsymbol{\alpha}$, $\boldsymbol{\beta}$, and $\boldsymbol{\gamma}$ levels, in the magnetic field range of interest ($B_{\perp} > 1$ T) where we observe the different ground states in sample C for $\nu<13/2$. As seen in Fig. \ref{fig:density}(e), $\boldsymbol{\alpha}$ has mainly two dominant spinor components: $N = 1$ and $N =3$. In contrast, $\boldsymbol{\beta}$ possesses predominantly an $N=1$ spinor component [Fig. \ref{fig:density}(f)], while $\boldsymbol{\gamma}$’s main components are $N=2$ and $N=4$. As we discuss below, the spinor components of $\boldsymbol{\alpha}$, $\boldsymbol{\beta}$, and $\boldsymbol{\gamma}$ can explain the evolution of the different states  at different fillings in Fig. \ref{fig:density}(a) to a large degree.

$\nu=$ \textit{5/2} --- In the density range we studied for sample C, $E_F$ at $\nu=5/2$ (1.9 T $<B_{\perp}<$ 3.6 T) remains in $\boldsymbol{\alpha}$ [Figs. (b,c,d)]. The dominating $N=1$ component is likely crucial for hosting the 5/2 FQHS, consistent with the fact that FQHSs at half fillings are predominantly observed in the $N=1$ LLs \cite{Willett.PRL.1987, Hossain.PRL.2018, Zibrov.Nature.2017, Li.Science.2017, Shi.NatNano.2020, Huang.PRX.2022}. The significant $N=3$ component ($>20\%$) at $B_{\perp}\geq2$ T, on the other hand, is very likely the origin of nematicity near $\nu=5/2$. We note that $\boldsymbol{\alpha}$ and $\boldsymbol{\beta}$ cross at very low $B_{\perp}$. Our calculations indicate that in the density range we measured for sample C, this crossing is reasonably far away from $E_F$, consistent with the absence of transitions between isotropic and anisotropic behaviors near $\nu=5/2$.

The nature of the nematic FQHS at $\nu=5/2$ in our 2DHS is not fully clear. While the 5/2 FQHS in the $N=1$ LL is generally associated with non-Abelian Pfaffian, anti-Pfaffian, or PH-Pfaffian topological orders \cite{Ma.preprint.2022}, the coexistence of nematic order with these topological orders requires further investigation. The theoretically proposed stripe FQHS \cite{Wan.PRB.2016} and pair-density-wave FQHS \cite{Santos.PRX.2019}, which can break the rotational symmetry at $\nu=5/2$, provide potential explanations for our observations.

The developing nematic FQHSs we observe at odd-denominator fillings flanking 5/2 also warrant discussion. They emerge at $\nu=8/3$, 13/5, 12/5, 7/3. Our LL calculations indicate that the LL ($\boldsymbol{\alpha}$) that hosts these states has predominantly $N=1$ component. Theories predict that the FQHSs at $\nu=12/5$ and 13/5 in the $N=1$ LL may support non-Abelian Fibonacci anyonic excitations, potentially useful for universal topological quantum computing \cite{Read.PRB.1999, Zhu.PRL.2015, Pakrouski.PRB.2016}. While a $\nu=12/5$ FQHS has been observed in the $N=1$ LL of GaAs 2DESs, the $\nu=13/5$ state has remained elusive \cite{Kumar.PRL.2010}. The absence of the $\nu=13/5$ FQHS was attributed to the effect of LL mixing \cite{Pakrouski.PRB.2016}. However, LL mixing is much larger in our 2DHS because of the larger hole effective mass, and yet our data reveal signatures of developing FQHSs at both $\nu=13/5$ and 12/5, with significant transport anisotropy; see Fig. \ref{fig:base}(c). In sample C, we also observe hints of developing FQHSs at $\nu=$ 12/5 and 13/5 at $p=2.18$; see SM Fig. S4 \cite{SM}. 

$\nu=$ \textit{7/2 and 9/2} --- The crossing between $\boldsymbol{\beta}$ and $\boldsymbol{\gamma}$ near $E_F$ and their spinor compositions can qualitatively explain the transitions between isotropic and anisotropic behaviors we observe in sample C near $\nu=7/2$ (1.3 T $<B_{\perp}<$ 2.6 T) and 9/2 (1.0 T $<B_{\perp}<$ 2.0 T). Thanks to the $N=2$ and $N=4$ components, anisotropic phases can be favored in $\boldsymbol{\gamma}$. In contrast, anisotropic phases are not favored in $\boldsymbol{\beta}$, because $\boldsymbol{\beta}$ does not possess any $N\geq2$ components. At low densities, $E_F$ for both $\nu=7/2$ and 9/2 is on the lower-$B_{\perp}$ side of the crossing. In this case, $\nu=7/2$ is in $\boldsymbol{\beta}$ while $\nu=9/2$ is in $\boldsymbol{\gamma}$. An isotropic phase near $\nu=7/2$ and an anisotropic phase near $\nu=9/2$ are expected. With increasing density, the crossing coincides with $E_F$ in the range of $3<\nu<5$, starting from the low-$\nu$ side. At intermediate densities when $E_F$ at $\nu=7/2$ has moved to the higher-$B_{\perp}$ side of the crossing while $E_F$ at $\nu=9/2$ is still at the lower-$B_{\perp}$ side of the crossing, both $\nu=7/2$ and 9/2 are in $\boldsymbol{\gamma}$, and anisotropic phases are expected near both $\nu=7/2$ and 9/2. At high densities, $E_F$ at both $\nu=7/2$ and 9/2 is on the higher-$B_{\perp}$ side of the crossing. In this case, $\nu=7/2$ is in $\boldsymbol{\gamma}$ while $\nu=9/2$ is in $\boldsymbol{\beta}$. An anisotropic phase near $\nu=7/2$ and an isotropic phase near $\nu=9/2$ are expected. These are qualitatively consistent with what we observe in sample C [Fig. \ref{fig:density}(a)].

We also note that the 2DHS (sample C) becomes slightly anisotropic near $\nu=9/2$ at $p>2.0$. However, an anisotropic behavior is not expected at high densities when $E_F$ is on the higher-$B_{\perp}$ side of the crossing between $\boldsymbol{\beta}$ and $\boldsymbol{\gamma}$. The anisotropy we observe at $p>2.0$ in sample C is possibly associated to a different crossing between $\boldsymbol{\beta}$ and another LL that has $N\geq2$ components.

$\nu=$ \textit{11/2, 13/2, 15/2, ...} --- Our calculations indicate that, in the density range we study for our samples, LLs at $\nu>5$ have predominantly $N\geq2$ components. This explains our observation of anisotropic phases at $\nu=$ 11/2, 13/2, 15/2, and 17/2 in sample A [Fig. \ref{fig:base}(a)] and at $p>2.0$ in sample C [Fig. \ref{fig:density}(a)]. At low densities (e.g., $p=1.13$), the absence of anisotropic phases at these fillings is likely because that such phases are very fragile and may require lower temperatures and disorder at lower densities.

It is worth noting that the agreement between our data and our LL calculations is qualitative. There are some discrepancies if we make quantitative comparisons. For example, in our LL calculations, we predict that $E_F$ at $\nu=7/2$ crosses from $\boldsymbol{\beta}$ to $\boldsymbol{\gamma}$ at $p\simeq1.0$, and that $E_F$ at $\nu=9/2$ crosses from $\boldsymbol{\gamma}$ to $\boldsymbol{\beta}$ at $p\simeq1.3$. In our experiment, however, the isotropic-to-anisotropic transition at $\nu=7/2$ occurs at $p\simeq1.3$, and the anisotropic-to-isotropic transition at $\nu=9/2$ occurs at $p\simeq1.9$. Similar quantitative discrepancies were also seen in several previous studies of crossings between other LLs in GaAs 2DHSs \cite{Liu.PRB.2014, Lupatini.PRL.2020, Ma.PRL.2022}. Although the origin of these discrepancies is unclear, it appears that the calculations always predict the crossings to occur at lower $B_{\perp}$ compared to experimental observations.

While our calculated LLs, including their compositions, capture some key experimental features, puzzles still remain. First, our LL calculations cannot explain the observed nematic FQHS at $\nu=7/2$. Our calculations indicate that the LL ($\boldsymbol{\gamma}$) that hosts the $\nu=7/2$ nematic FQHS only has $N\geq2$ components, but does not have any $N=1$ component. This is consistent with the absence of FQHSs at odd-denominator fillings that may be favored in an $N=1$ LL. The emergence of a nematic FQHS at $\nu=7/2$ in a $N=2$-dominant LL suggests a distinct origin compared to the nematic FQHS at $\nu=5/2$. Another puzzling observation relates to the temperature dependence of the anisotropic phases at $\nu=5/2$ and 7/2 in tilted $B$. In the $N\geq2$ LLs of GaAs 2DESs, the anisotropic phases typically become more robust in tilted $B$ as compared to in purely $B_{\perp}$ \cite{Cooper.PRB.2002}. Unexpectedly, we find that at $\theta=66.2^{\circ}$, the anisotropic phases at $\nu=5/2$ and 7/2 turn isotropic at much lower $T$ [$\simeq75$ mK; see Fig. S7 in SM \cite{SM}] compared to what we observe in purely $B_{\perp}$ [$\gtrsim$ 150 mK; see Figs. \ref{fig:Tdep}(g,h)].

A phase transition from an isotropic CF Fermi sea to a highly anisotropic FQHS was reported at $\nu=3/2$ in GaAs 2DHSs under tilted $B$ \cite{Wang.PRL.2023}. While such an anisotropic FQHS is likely also associated with a LL possessing multiple spinor components, there are some key differences: (i) The large in-plane $B$ in tilted $B$ explicitly breaks the rotational symmetry and favors anisotropic phases at $\nu=3/2$, in stark contrast to the spontaneous rotational symmetry breaking we observe here at $\nu=5/2$ and 7/2. (ii) Our observation of nematic FQHSs at $\nu=5/2$ and 7/2, as well as the phase transitions at $\nu=7/2$ and 9/2 show clear correlation with the calculated LL spinor components and crossings. However, similar calculations for LLs and their spinor components in tilted $B$ are not available because of computational challenges. This hinders a quantitative LL analysis of the anisotropic FQHS at $\nu=3/2$ which was only seen in tilted $B$.

\section{Conclusion}

Our observation of nematic FQHSs with spontaneously broken rotational symmetry at both even- and odd-denominator $\nu$, as well as the transitions between isotropic and anisotropic phases, highlight the rich physics in the quantum Hall regime of ultrahigh-quality GaAs 2DHSs. The effects of heavy-hole light-hole and spin-orbit couplings render GaAs hole LLs multi-orbital, making them unique systems that support coexisting topological and nematic orders. Furthermore, the tunability of LL components and crossings provide useful platforms for realizing and engineering novel, exotic, interaction phenomena, and studying phase transitions between distinct strongly correlated phases.

Our work also highlights the crucial role of sample quality in the exploration of emergent many-body phenomena. Improvements in atomically-thin 2D materials, such as graphene-based systems and transition metal dichalcogenides, have also made available new types of material platforms for studies of FQHSs and other many-body phases \cite{Ki.NanoLett.2014, Zibrov.Nature.2017, Li.Science.2017, Huang.PRX.2022, Shi.NatNano.2020}. These new materials have significantly expanded the scope of FQHS research by introducing different tuning knobs and probing techniques, enabling exciting observations \cite{Tsui.Nature.2024, Hu.NatPhys.2025}. Among a variety of materials, the modulation-doped GaAs/AlAs quantum wells yet remain a leading platform for the discovery of new emergent phenomena and for potential applications, thanks to their record-high transport mobility \cite{Chung.Natmater.2021, Chung.PRM.2022, Gupta.PRM.2024}. An additional key advantage of the GaAs/AlAs system is the large sample size ($\sim$ 25 mm$^2$), about 10$^6$ times larger than devices made of exfoliated 2D flakes.

\section*{Acknowledgement} 
We acknowledge support by the National Science Foundation (NSF) Grant No. DMR 2104771 for measurements, and the Gordon and Betty Moore Foundation’s EPiQS Initiative (Grant No. GBMF9615 to L.N.P.) for sample fabrication. Our measurements were partly performed at the National High Magnetic Field Laboratory (NHMFL), which is supported by the NSF Cooperative Agreement No. DMR 2128556, by the State of Florida, and by the DOE. This research was funded in part by QuantEmX grant from Institute for Complex Adaptive Matter and the Gordon and Betty Moore Foundation through Grant GBMF9616 to C. W., A. G., S. K. S., and C. T. T. We thank A. Bangura, G. Jones, R. Nowell and T. Murphy at NHMFL for technical assistance, and J. K. Jain for illuminating discussions.



\section*{Supplementary material}

# Supplemental Material to "Even-denominator fractional quantum Hall states with spontaneously broken rotational symmetry"

Chengyu Wang,[1] A. Gupta,[1] S. K. Singh,[1] C. T. Tai,[1] L. N. Pfeiffer,[1] K. W. Baldwin,[1] R. Winkler,[2] and M. Shayegan[1]

[1]Department of Electrical and Computer Engineering,
Princeton University, Princeton, New Jersey 08544, USA
[2]Department of Physics, Northern Illinois University, DeKalb, Illinois 60115, USA
(Dated: September 22, 2025)

## I. SAMPLE PARAMETERS

Table I summarize the sample parameters for the ultrahigh-quality GaAs two-dimensional hole systems (2DHSs) we used in this work. We studied seven samples from six different wafers. The 2DHSs are confined to relatively narrow GaAs quantum wells (QWs), and only have the ground-state electric subband occupied. Samples A and B are from the same wafer, and have a density ($p$) of 2.1 in units of $10^{11}$ cm$^{-2}$, which we use throughout the manuscript. Sample C has an as-grown density of 1.7. This sample was fitted with a Ti/Au front gate and an In back gate, allowing for *in-situ* tuning of the 2D hole density. All the samples we study have transport mobility higher than $1.5 \times 10^6$ cm$^2$/Vs.

## II. ADDITIONAL MAGNETO-TRANSPORT DATA FOR SAMPLE A

Figures S1 (a, b) show $R_{xx}$ and $R_{yy}$ data for sample A near $\nu = 7/2$ and $5/2$, respectively. In panel (c), we show schematics of the circuit geometries we use for $R_{xx}$ and $R_{yy}$ measurements. We observe clear $R_{xx}$ and $R_{yy}$ minima at even-denominator fillings $\nu = 7/2$ and $5/2$, and odd-denominator fillings $\nu = 14/5$, $8/3$, $13/5$, $12/5$, $7/3$, and $11/5$, consistent with Fig. 2 in the main text. The observation of minima in both $R_{xx}$ and $R_{yy}$ further supports the notion of developing fractional quantum Hall states (FQHSs) at these fillings. We also observe pronounced minima in both $R_{xx}$ and $R_{yy}$ at $B_\perp \simeq 2.63$ T, consistent with the notion of a developing reentrant integer quantum Hall state, as well as minima at $B_\perp \simeq 2.68$ T ($\nu = 16/5$), suggesting a developing FQHS at this fill-

ing. Note that $R_{xx}$ in Fig. 2 of the main text approaches zero in a range of $\nu$ near $\nu = 7/2$ and $5/2$ where the 2DHS is anisotropic. This is because the data in Fig. 2 of the main text were measured using the four corner contacts, and the relatively large distance from the voltage probes to the source and drain cause a significant decay of current density near the voltage probes [1]. With the circuit geometry shown in Fig. S1(c), $R_{xx}$ exhibits finite resistance when the 2DHS is anisotropic ($R_{xx} < R_{yy}$).

The transport energy gap ($\Delta$) of a FQHS is typically deduced from the relation $R_{xx} \propto e^{-\Delta/2k_BT}$. For our 2DHS, however, the temperature dependence of $R_{xx}$ and $R_{yy}$ is dominated by the anisotropic resistance background; see Fig. 3 of the main text. A useful method to estimate the energy gap of a FQHS riding on a temperature-dependent resistance background is to extract a so-called pseudogap from the temperature dependence of the relative depth of the minimum [2]. In Fig. S2, we present Arrhenius plots of $R_{5/2}/R_{bg}$ and $R_{7/2}/R_{bg}$ vs $1/T$, where $R_{bg}$ are the background resistances defined in insets of Fig. S2. Slopes of the linear fits give estimates of energy gaps for FQHSs at $\nu = 5/2$ ($\Delta \simeq 55$ mK) and 7/2 ($\Delta \simeq 78$ mK). The small energy gaps for the $\nu = 5/2$ and 7/2 FQHSs are consistent with the fact that these FQHSs are only observed at very low temperatures ($T < 50$ mK).

In Fig. S3, for sample A, we present Hall resistance $R_H$ near $\nu = 3/2$ measured at $T \simeq 20$ mK, and $R_{xx}$ and $R_{yy}$ data near $\nu = 3/2$ measured at different temperatures. Near $\nu = 3/2$, both $R_{xx}$ and $R_{yy}$ are smooth and nearly temperature independent. At base temperature $T \simeq 20$ mK, $R_H$ is linear as a function of $B_\perp$ near $\nu = 3/2$. On the flanks of $\nu = 3/2$, we observe FQHSs at $\nu = 1 + n/(2n \pm 1)$, where $n = 1, 2, 3, 4$. These observations are consistent with an isotropic composite fermion (CF) Fermi sea at the half filling $\nu = 3/2$ flanked by the standard Jain-sequence FQHSs, which are expected in an $N = 0$ Landau level (LL).

## III. DATA FOR OTHER SAMPLES

The signatures of nematic, even-denominator, FQHSs at $\nu = 5/2$ and 7/2 are reproducible in multiple samples with different parameters (e.g., density and QW width). In this section, we present magneto-transport data near $\nu = 5/2$ and 7/2 for samples C-F.

Figure S4(a) shows a Hall trace for sample C mea-

TABLE I. **Sample parameters**

| Sample | density ($10^{11}$ cm$^{-2}$) | mobility ($10^6$ cm$^2$/Vs) | QW width (nm) | Al fraction for the barrier near the QW |
|---|---|---|---|---|
| A | 2.1 | 2.1 | 20 | 0.08 |
| B | 2.1 | 2.0 | 20 | 0.08 |
| C | 1.7 | 3.6 | 20 | 0.16 |
| D | 2.6 | 1.5 | 20 | 0.32 |
| E | 2.1 | 2.6 | 15.75 | 0.08 |
| F | 2.1 | 2.1 | 17.5 | 0.08 |
| G | 2.2 | 1.8 | 20 | 0.26 |

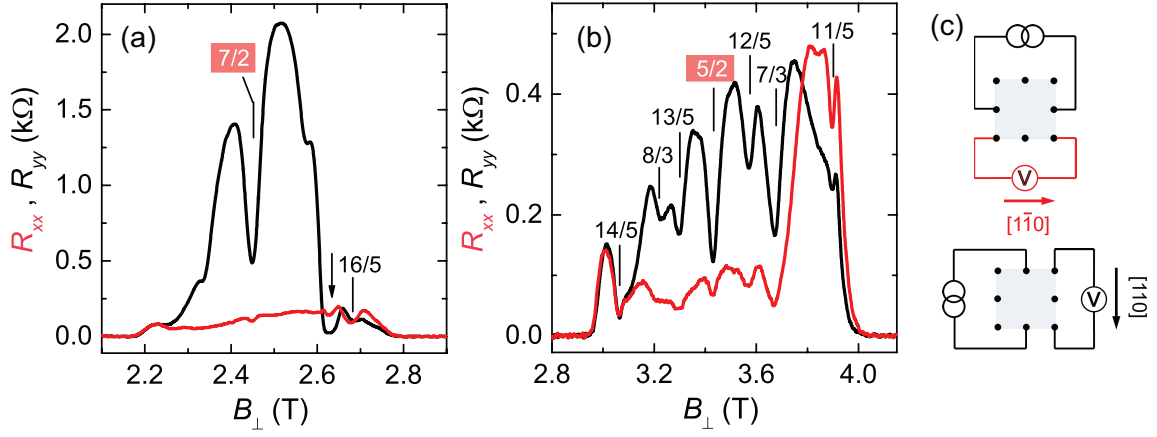

Fig. S1. **Data for sample A measured at $T \simeq 20$ mK using different circuit configurations.** (a,b) $R_{xx}$ (along [1$\bar{1}$0]) and $R_{yy}$ (along [110]) data near $\nu = 7/2$ and $5/2$, measured using circuit configurations shown in panel (c) on the right. Clear minima are observed in both $R_{xx}$ and $R_{yy}$ at even-denominator fillings $\nu = 7/2$, $5/2$, and odd-denominator fillings 16/5, 14/5, 8/3, 13/5, 12/5, 7/3, and 11/5, superimposed on an anisotropic resistance background.

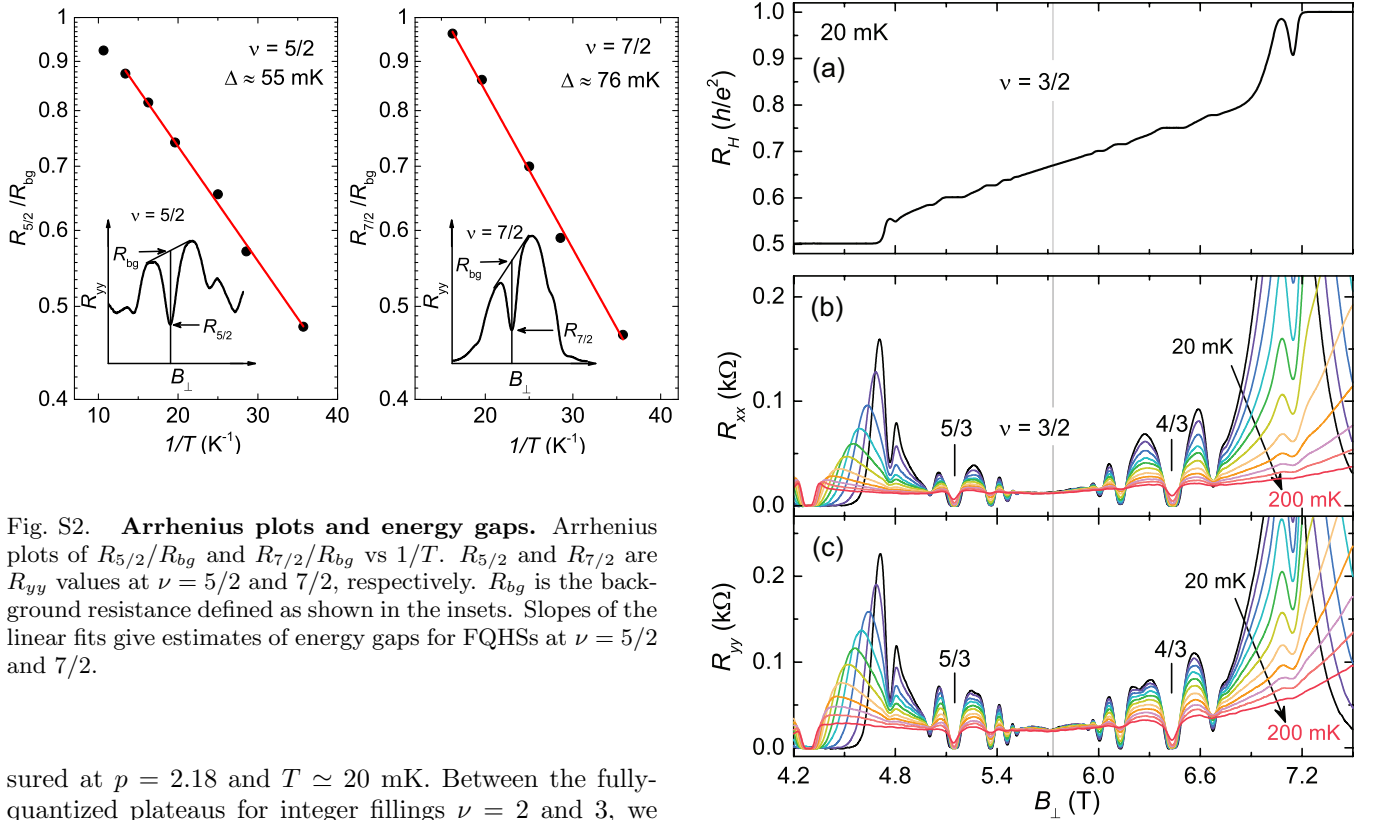

Fig. S2. **Arrhenius plots and energy gaps.** Arrhenius plots of $R_{5/2}/R_{bg}$ and $R_{7/2}/R_{bg}$ vs $1/T$. $R_{5/2}$ and $R_{7/2}$ are $R_{yy}$ values at $\nu = 5/2$ and $7/2$, respectively. $R_{bg}$ is the background resistance defined as shown in the insets. Slopes of the linear fits give estimates of energy gaps for FQHSs at $\nu = 5/2$ and $7/2$.

Fig. S3. **Data near $\nu = 3/2$.** (a) Hall resistance $R_H$ vs $B_\perp$ measured at $T \simeq 20$ mK. (b, c) Longitudinal resistances $R_{xx}$ and $R_{yy}$ vs $B_\perp$ near $\nu = 3/2$ measured at different temperatures. The $R_{xx}$ and $R_{yy}$ values near $\nu = 3/2$ are nearly independent on temperature.

sured at $p = 2.18$ and $T \simeq 20$ mK. Between the fully-quantized plateaus for integer fillings $\nu = 2$ and 3, we observe developing plateaus at the even-denominator filling $\nu = 5/2$ and at odd-denominator fillings $\nu = 14/5$, 8/3, 13/5, 12/5, and 7/3. We note that the developing plateaus at $\nu = 12/5$ and 7/3 are about 1% higher than their expected quantized $R_H$ values. In Fig. S4(b), we present Hall slope $dR_H/dB_\perp$, $R_{xx}$, and $R_{yy}$ vs $B_\perp$. We observe clear $dR_H/dB_\perp$ minima at $\nu = 5/2$, 14/5, 8/3, 13/5, 12/5, and 7/3. The 2DHS is anisotropic near $\nu = 5/2$, with $R_{yy} \gg R_{xx}$. Clear $R_{yy}$ minima are seen at $\nu = 5/2$ and 7/3, and weaker $R_{yy}$ minima/inflection

points are identified at $\nu = 14/5$, 8/3, 13/5, 12/5, and

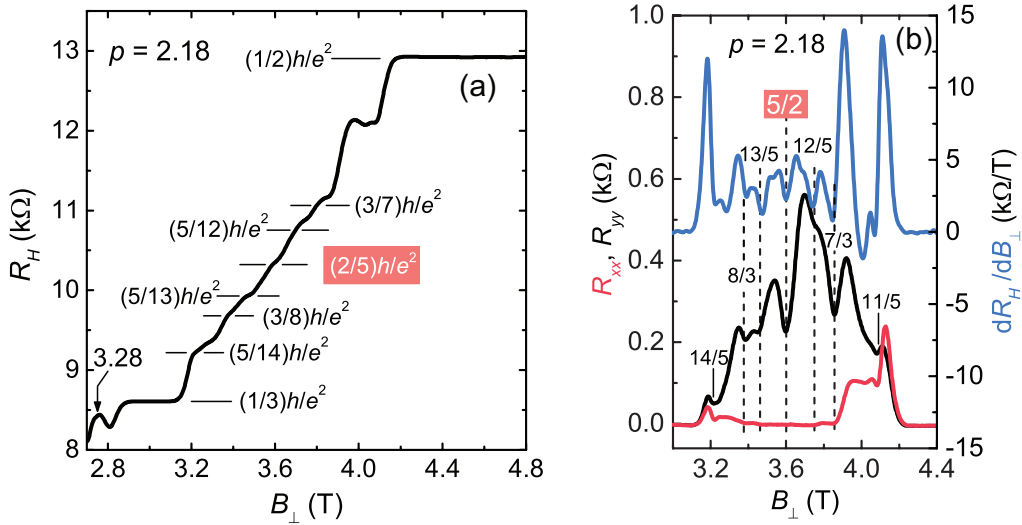

Fig. S4. **Hall data for sample C.** (a) Hall resistance $R_H$ vs $B_\perp$ trace, measured at $p = 2.18$ and $T \simeq 20$ mK. Between the integer quantized plateaus for $\nu = 2$ and 3, we observe developing plateaus at even-denominator filling $\nu = 5/2$, as well as at odd-denominator fillings $\nu = 14/5$, 8/3, 13/5, 12/5, and 7/3. (b) $R_{xx}$, $R_{yy}$, and $dR_H/dB_\perp$ traces, measured at $p = 2.18$ and $T \simeq 20$ mK.

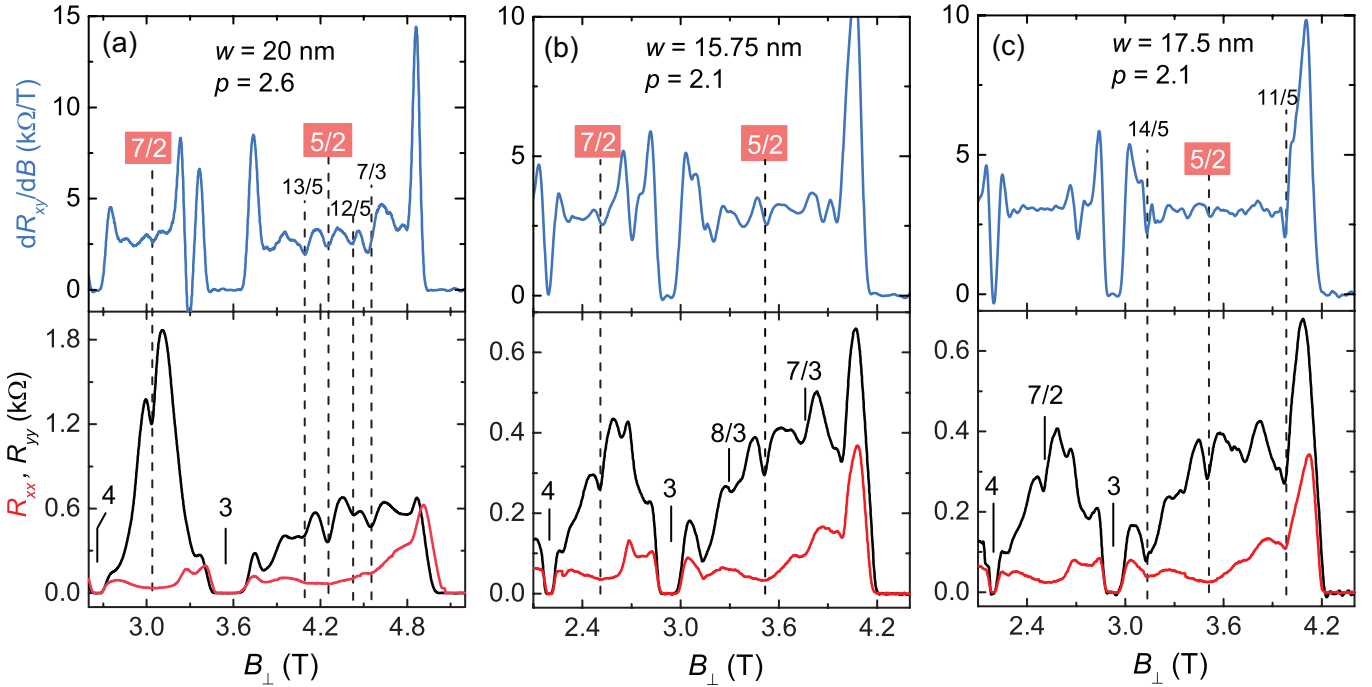

Fig. S5. **Signatures of nematic FQHSs in different samples.** $R_{xx}$, $R_{yy}$ and $dR_H/dB_\perp$ data near $\nu = 5/2$ and 7/2 for samples D, E, and F, measured at $T \simeq 30$ mK. (a) Sample D, $w = 20$ nm, $p = 2.6$, (b) Sample E, $w = 15.75$ nm, $p = 2.1$, (c) Sample F, $w = 17.5$ nm, $p = 2.1$.

11/5. Our observations are consistent with what we find in sample A; see Fig. 2(c) of the main text.

Figure S5 shows $R_{xx}$, $R_{yy}$, and the Hall slope $dR_H/dB_\perp$ data for samples D, E and F. The circuit configurations used for the $R_{xx}$ and $R_{yy}$ measurements are

the same as what we show in Fig. S1(c). Results qualitatively similar to what we observe in samples A, B, and C are also found in samples D, E, and F: The 2DHSs are anisotropic near $\nu = 5/2$ and 7/2, with $R_{yy} >> R_{xx}$. We observe sharp minima in $R_{yy}$ and broader minima

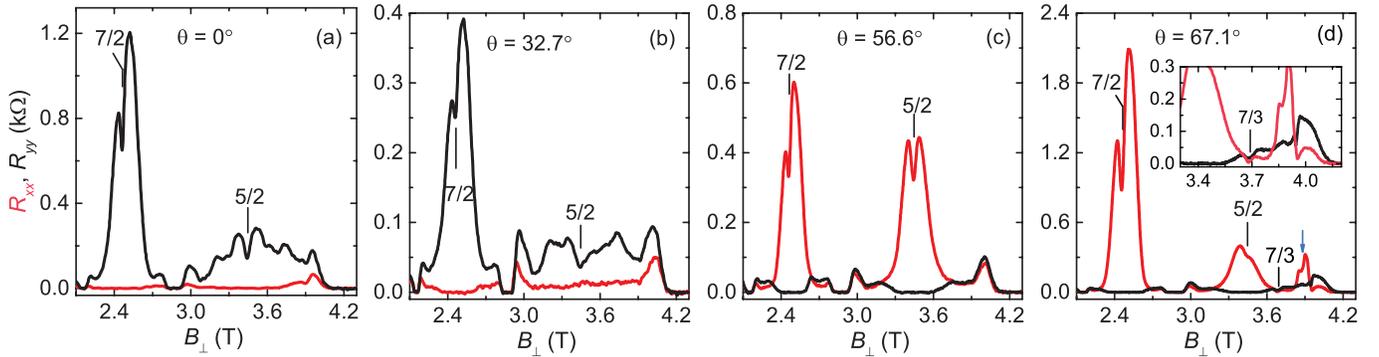

Fig. S6. **Tilted magnetic field data for sample B at** $T \simeq 30$ mK. (a) $R_{yy}$ vs $B_\perp$ traces measured at different tilt angles. The tilt angle $\theta$ for each trace is shown on the right side. (b-d) $R_{xx}$ and $R_{yy}$ vs $B_\perp$ at different tilt angles (b) $\theta = 0°$, (c) $\theta = 32.7°$, (c) $\theta = 56.6°$, (d) $\theta = 67.1°$. Inset in (d) shows $R_{xx}$ and $R_{yy}$ near $\nu = 7/3$ on an expanded scale. For the sample geometry and definition of $\theta$, see the inset to Fig. 4(a) in the main text.

in $R_{xx}$ at $\nu = 5/2$ and $7/2$. We also observe minima in $dR_H/dB_\perp$ at $\nu = 5/2$ and $7/2$, indicating developing Hall plateaus at these fillings. These observations confirm developing, nematic FQHSs at $\nu = 5/2$ and $7/2$.

## IV. ADDITIONAL DATA IN TILTED MAGNETIC FIELD

In this Section, we present additional data for samples B and G measured in tilted $B$.

As we show in Figs. 4(c,d) of the main text, a large in-plane magnetic field applied along [1$\bar{1}$0] direction switches the high resistance direction (hard axis) of the $\nu = 5/2$ and $7/2$ states from [110] to [1$\bar{1}$0]. In Fig. S6, we present $R_{xx}$ and $R_{yy}$ data for sample B near $\nu = 5/2$ and $7/2$ at four different tilt angles ($\theta$) to show the evolution of different states with increasing $\theta$.

At $\theta = 0$, we observe developing FQHSs at $\nu = 5/2$ and $7/2$ with $R_{xx} << R_{yy}$. At $\theta = 32.7°$, the $R_{yy}$ minima at $\nu = 5/2$ and $7/2$ become shallower and the anisotropy near $\nu = 5/2$ and $7/2$ also becomes smaller. At larger $\theta = 56.6°$, $R_{xx}$ becomes larger than $R_{yy}$ near $\nu = 5/2$ and $7/2$ [Fig. S6(c)], suggesting that developing FQHSs at $\nu = 5/2$ and $7/2$ survive at large $\theta$. In contrast, no FQHS feature is seen at any odd-denominator $\nu$ near $\nu = 5/2$ and $7/2$.

Several observations are made at an even higher tilt angle $\theta = 67.1°$ [Fig. S6(d)]: (i) While the anisotropy near $\nu = 5/2$ remains similar to that at $\theta = 56.6°$, the $R_{xx}$ minimum at $\nu = 5/2$ disappears. (ii) We observe a nearly isotropic FQHS at $\nu = 7/3$, evinced by in a sharp minimum in both $R_{xx}$ and $R_{yy}$ [see inset in Fig. S6(d)]. (iii) We observe a sharp feature at $B_\perp \simeq 3.9$ T, where $R_{xx}$ suddenly becomes larger than $R_{yy}$ [see blue arrow in Fig. S6(d)]. While the origin of this anisotropic state is unclear, we speculate that it is likely associated with a LL crossing induced by large $\theta$. (iv) Compared to the

data at $\theta = 56.6°$, the anisotropy near $\nu = 7/2$ becomes larger and the $R_{xx}$ minimum at $\nu = 7/2$ becomes more pronounced. We also note that at large $\theta$, the $\nu$ range for the anisotropic $R_{xx}$ and $R_{yy}$ near $\nu = 5/2$ is narrower compared to when $\theta = 0°$. At both $\theta = 0°$ and the largest $\theta = 67.1°$, the anisotropy at $\nu = 7/2$ is larger than that at $\nu = 5/2$.

It is worth noting that large in-plane magnetic field can potentially also modify the LL structure of 2DHSs because of the spin-orbit coupling and heavy-hole light-hole interaction. Such an effect was in fact utilized to induce a crossing between two LLs with different spinor components [3, 4]. On the other hand, our observation in sample B near $\nu = 5/2$ and $7/2$ at large $\theta$, i.e., that the hard axis of anisotropy aligns with the in-plane magnetic field direction, is consistent with Hartree-Fock calculations in Ref. [5]. Therefore, we do not need to invoke changes in the LL structure to interpret our observation. Interestingly, the anomalous feature we observe near $B_\perp \simeq 3.9$ T at the largest $\theta$ (67.1°) is likely associated with a LL crossing induced by large in-plane $B$; see Fig. S6(d). A quantitative understanding of the LL structure of 2DHSs in tilted $B$ is by itself a very interesting topic. Unfortunately, however, calculations for LLs and their spinor components in tilted $B$ are not available because of computational challenges.

Figure S7 shows the temperature dependence data at a large tilt angle of 66.2°. We observe anisotropic phases at $\nu = 5/2$, $7/2$, and $9/2$, evinced by the opposite trends in $R_{xx}$ and $R_{yy}$ as a function of temperature. At $T \simeq 30$ mK, weak anisotropic phases are also seen at $\nu = 11/2$ and $13/2$ where $R_{xx}$ exhibits a maximum and $R_{yy}$ exhibits a minimum. Figures S7(c-e) show $R_{xx}$ and $R_{yy}$ at $\nu = 5/2$, $7/2$, and $9/2$ as a function of $T$. We note that the anisotropy at $\nu = 5/2$ and $7/2$ starts to develop below $T \simeq 75$ mK, lower than what we observe at $\theta = 0$ ($T \simeq 150$ mK).

In Fig. S8, we present $R_{xx}$ and $R_{yy}$ data for sample G measured at $T \simeq 30$ mK and at different $\theta$. Compared to

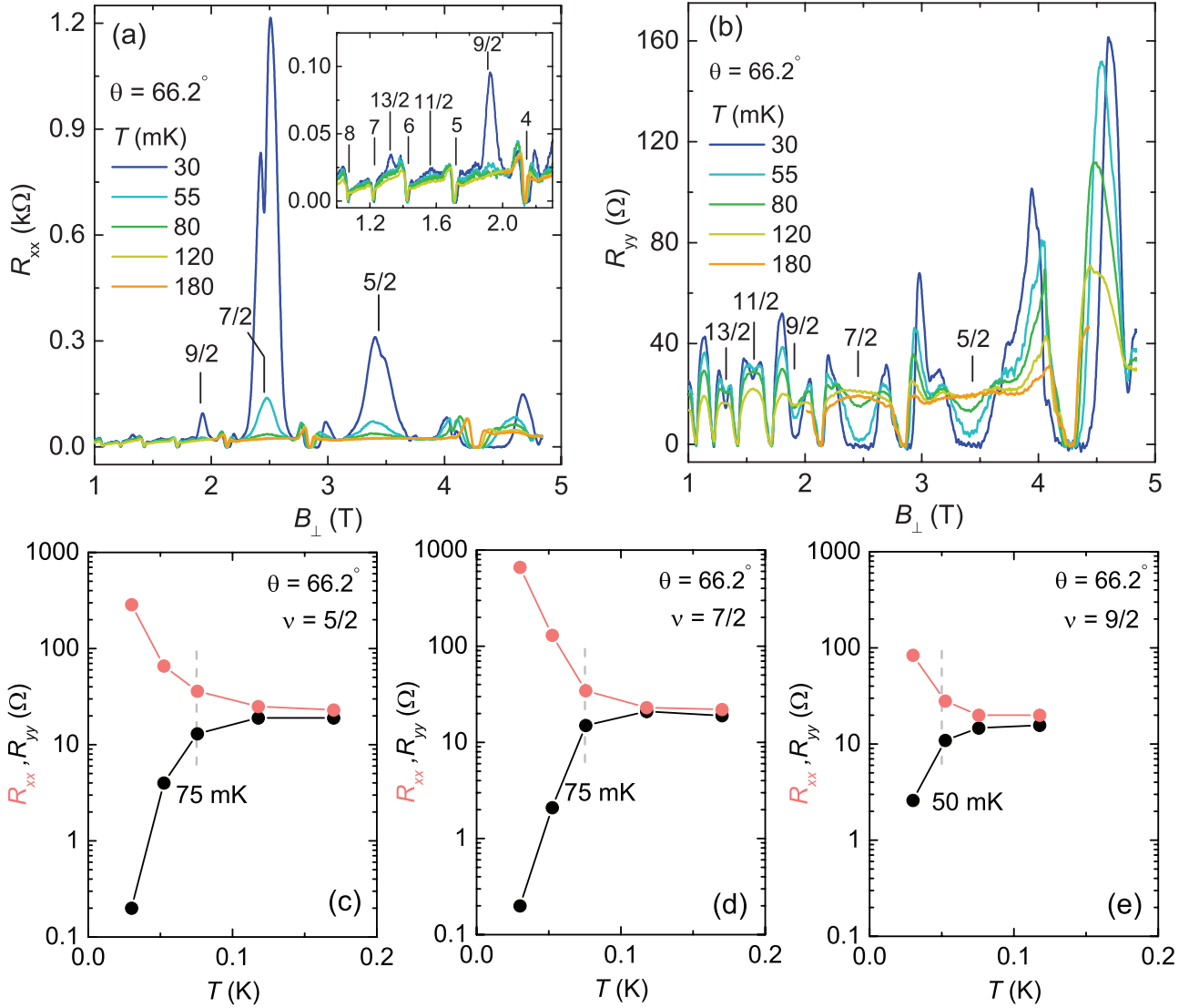

Fig. S7. **Temperature dependence of tilted-field data at θ = 66.2° for sample B.** (a, b) $R_{xx}$ and $R_{yy}$ vs $B_\perp$ measured at different temperatures. Inset in (a) shows an expanded version of data at high fillings. (c-e) $R_{xx}$ and $R_{yy}$ vs $T$ at $\nu = 5/2$, 7/2, and 9/2.

samples A and B we show in the main text, this sample has the same QW width (20 nm), a slightly higher 2D hole density of 2.2, and a much higher Al content of the AlGaAs barrier near the QW ($x = 0.26$). At $\theta = 0$, the 2DHS is anisotropic near $\nu = 5/2$ and 7/2 with $R_{xx} \gg R_{yy}$. Sharp minima in $R_{yy}$ are seen at $\nu = 5/2$ and 7/2. At $\theta = 41.5°$, the anisotropy near $\nu = 7/2$ becomes smaller and the $R_{xx}$ and $R_{yy}$ values become close near $\nu = 5/2$. At $\theta = 60.5°$, the 2DHS becomes anisotropic at both $\nu = 5/2$ and 7/2, but now with $R_{xx} \gg R_{yy}$. The interchange of hard- and easy-axis directions at large $\theta$ is consistent with our observations in sample B; see Fig. 4 in the main text.

## V. LANDAU LEVEL CALCULATIONS FOR GaAs 2D HOLES

In the main text, we show the results of LL calculations for our 2DHS. Here we present a more detailed discussion of the calculations of GaAs hole LLs in $B_\perp$.

Holes in the topmost valence band have an effective spin 3/2. In a QW, the four-fold degeneracy is lifted, and we have heavy holes (HHs) corresponding to spin $z$ component $s = \pm 3/2$ and light holes (LHs) with $s = \pm 1/2$. For in-plane wave vector $k_\parallel = 0$ and $B_\perp = 0$ the eigenstates are purely HH or LH, whereas $k_\parallel \neq 0$ or $B_\perp \neq 0$ give rise to HH-LH coupling. (For conceptual clarity we restrict the discussion to the so-called axial approximation for 2D holes. We also focus on the

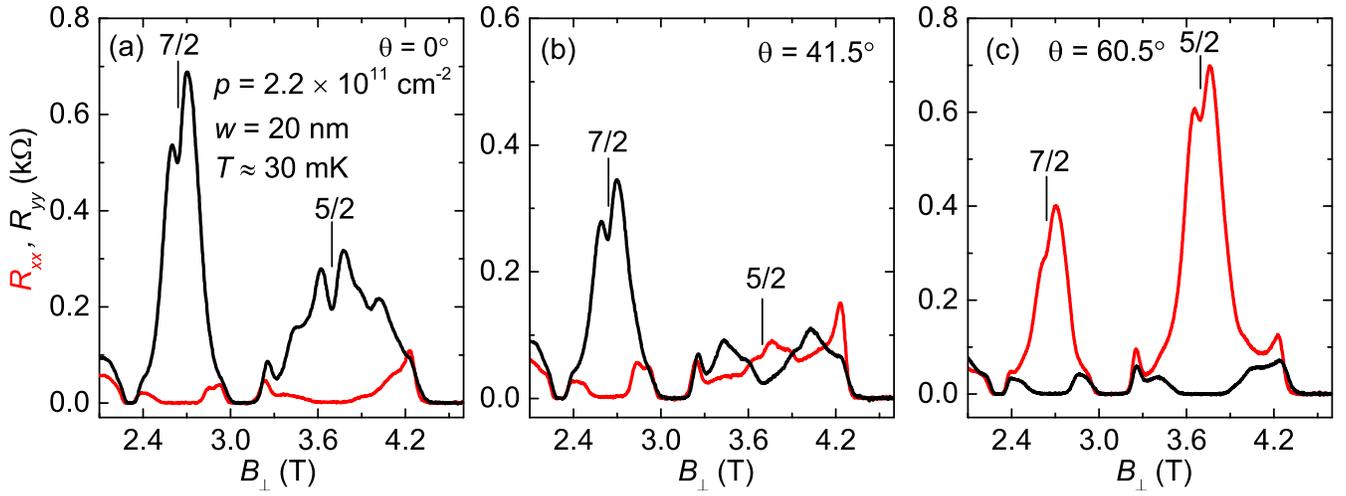

Fig. S8. **Tilt data for sample G at** $T \simeq$ **30 mK.** $R_{xx}$ and $R_{yy}$ data for sample G at three different tilt angles (a) $\theta = 0$, (b) $\theta = 41.5°$, (c) $\theta = 60.5°$.

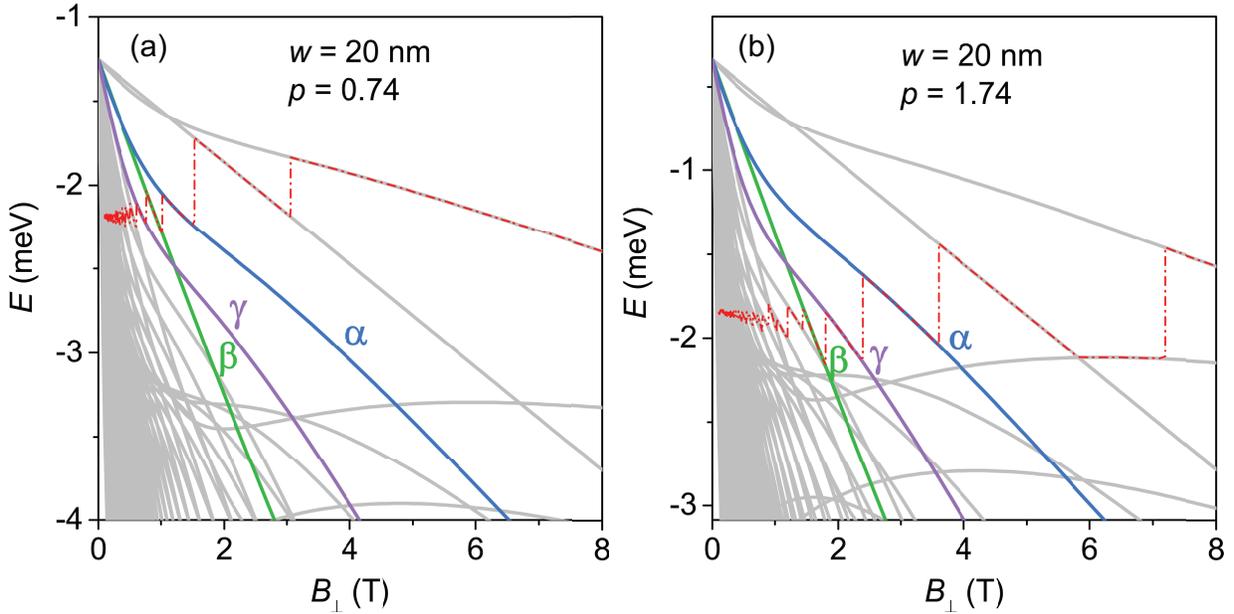

Fig. S9. **Calculated Landau levels for sample C at two different densities.** Calculated $E$ vs $B_\perp$ LL fan diagrams for sample C at (a) $p = 0.74$, (b) $p = 1.74$. Most relevant LLs (the $\boldsymbol{\alpha}$, $\boldsymbol{\beta}$, and $\boldsymbol{\gamma}$ levels) are highlighted. The red dash-dotted lines trace the Fermi energy $E_F$.

subspace representing the topmost valence band, and ignore excited subbands. A more complete discussion can be found in Ref. [6].) The eigenfunctions of the Hamiltonian are then four-component spinors representing $s = +3/2, +1/2, -1/2$ and $-3/2$.

The LLs in 2D systems can be described using two sets

of dimensionless ladder operators [6–8]:

$$a = \frac{-i}{\sqrt{2}}\left(\frac{\bar{w}}{2\ell} + 2\ell\frac{\partial}{\partial w}\right), \quad a^\dagger = \frac{i}{\sqrt{2}}\left(\frac{w}{2\ell} - 2\ell\frac{\partial}{\partial \bar{w}}\right),$$
(1a)

$$b = \frac{1}{\sqrt{2}}\left(\frac{w}{2\ell} + 2\ell\frac{\partial}{\partial \bar{w}}\right), \quad b^\dagger = \frac{1}{\sqrt{2}}\left(\frac{\bar{w}}{2\ell} - 2\ell\frac{\partial}{\partial w}\right),$$
(1b)

corresponding to two harmonic oscillators. Here $w = x + iy$, the bar indicates complex conjugation, and $\ell =$

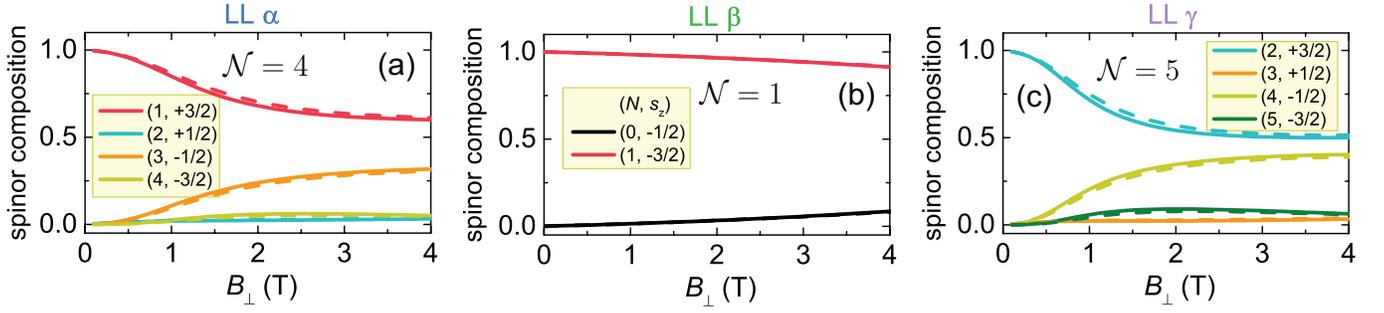

Fig. S10. **Spinor components.** (a-c) Calculated spinor components as a function of $B_\perp$ for $\boldsymbol{\alpha}$, $\boldsymbol{\beta}$, and $\boldsymbol{\gamma}$ levels. Soild lines are calculations for $p = 1.74$, dashed lines are calculations for $p = 0.74$.

$\sqrt{\hbar/(eB_\perp)}$ is the magnetic length. The single-particle Hamiltonian $H$ only depends on the $a$ oscillators; the $a$ oscillators thus define the energy spectrum of $H$ as a function of $B_\perp$. The $b$ oscillators commute with $H$, and represent an oscillator with frequency zero that describes the degeneracy of the LLs.

When $B_\perp \neq 0$, the total angular momentum perpendicular to the 2D plane $j = l + s$ remains a good quantum number, where:

$$l = xp_y - yp_x = \hbar \left( a^\dagger a - b^\dagger b \right) \qquad (2)$$

is the orbital angular momentum [6, 7]. The LLs with quantum number $\mathcal{N} = 0, 1, 2, \ldots$ can thus be written as [6]:

$$\psi_{\mathcal{N},M}(\boldsymbol{r}) = \sum_s |N = \mathcal{N} - s - \tfrac{3}{2}\rangle \, |M\rangle \, \xi_s^\mathcal{N}(z) \, u_s(\boldsymbol{r}). \qquad (3)$$

Here $|N\rangle$ with $N = 0, 1, 2, \ldots$ denotes the eigenstates of the $a$ oscillator with $|N\rangle = 0$ when $N < 0$, $|M\rangle$ with $M = 0, 1, 2, \ldots$ are the eigenstates of the $b$ oscillator, $\xi_s^\mathcal{N}(z)$ expresses the weight of each spinor component $s$, and $u_s(\boldsymbol{r})$ are band-edge Bloch functions that act as the basis vectors for the spinor components of the multispinor wave functions $\psi_{\mathcal{N},M}(\boldsymbol{r})$.

The index $N$ representing the states $|N\rangle$ is an important quantum number in the fractional quantum Hall regime. In 2DESs, each LL is characterized by one value of $N$, and LLs with different $N$ host different ground states. According to Eq. (3), the LLs formed by 2D holes are linear combinations of different Landau oscillators $|N\rangle$ representing the four spinor components.

We focus our LL calculations on sample C whose 2D hole density we can tune in our experiments. In Fig. S9, we present the calculated $E$ vs. $B_\perp$ LL diagrams for sample C at two different 2D hole densities. Because of the HH-LH coupling, LLs are highly nonlinear as a function of $B_\perp$ and show numerous crossings and anticrossings. The most relevant hole LLs in our study are the $\boldsymbol{\alpha}$, $\boldsymbol{\beta}$, and $\boldsymbol{\gamma}$ levels, which are highlighted in Fig. S9. Within the density range $0.74 < p < 1.74$, the Fermi energy $E_F$ for $2 < \nu < 3$ is located at the $\boldsymbol{\alpha}$ level, and $E_F$ for $3 < \nu < 5$ are located at the $\boldsymbol{\beta}$ and $\boldsymbol{\gamma}$ levels. We note that there is a crossing between the $\boldsymbol{\beta}$ and $\boldsymbol{\gamma}$ levels at $B_\perp \simeq 1.2$ T, which can coincidence with $E_F$ for $3 < \nu < 5$.

Figure S10 shows the spinor compositions as a function of $B_\perp$ for $\boldsymbol{\alpha}$, $\boldsymbol{\beta}$, and $\boldsymbol{\gamma}$ levels, calculated at two different densities ($p = 0.74$ and 1.74). We note that density has a minor effect on the $B_\perp$ dependence of the spinor compositions for these LLs. The spinor compositions for the $\boldsymbol{\alpha}$, $\boldsymbol{\beta}$, and $\boldsymbol{\gamma}$ levels are very different. The $\boldsymbol{\alpha}$ level has a quantum number $\mathcal{N} = 4$. It can be decomposed into the basis of four spinors with $(N, s_z) = (1, +3/2)$, $(2, +1/2)$, $(3, -1/2)$, and $(4, -3/2)$. Among these spinors, the $N = 1$ component is dominant ($> 60\%$) and the $N = 3$ component is significant ($> 20\%$) at $B_\perp > 2$ T. The $\boldsymbol{\beta}$ level has a quantum number $\mathcal{N} = 1$. It only has two spinor components with $(N, s_z) = (0, -1/2)$ and $(1, -3/2)$. The $N = 1$ component is dominant while the $N = 0$ component is negligible. The $\boldsymbol{\gamma}$ level has a quantum number $\mathcal{N} = 5$. It can be decomposed into the basis of four spinors with $(N, s_z) = (2, +3/2)$, $(3, +1/2)$, $(4, -1/2)$, and $(5, -3/2)$. The $N = 2$ component is dominant ($> 50\%$) and the $N = 4$ component is significant ($> 30\%$) at $B_\perp > 1.5$ T. It is worth noting that the $N \geq 2$ components possessed by the $\boldsymbol{\alpha}$ and $\boldsymbol{\gamma}$ levels are likely crucial for hosting anisotropic phases near half fillings. The $\boldsymbol{\beta}$ level, on the other hand, does not have any $N \geq 2$ components. Anisotropic phases are therefore not favored in the $\boldsymbol{\beta}$ level.

[1] Steven H. Simon, Comment on "Evidence for an Anisotropic State of Two-Dimensional Electrons in High Landau Levels", Phys. Rev. Lett. **83**, 4223 (1999).

[2] J. R. Mallett, R. G. Clark, R. J. Nicholas, R. Willett, J. J.




\begin{thebibliography}{1}

\bibitem{Tsui.PRL.1982} D. C. Tsui, H. L. Stormer, and A. C. Gossard, Two-Dimensional Magnetotransport in the Extreme Quantum Limit, Phys. Rev. Lett. {\bf 48}, 1559 (1982).

\bibitem{Jain.PRL.1989} J. K. Jain, Composite-fermion approach for the fractional quantum Hall effect, Phys. Rev. Lett. {\bf 63}, 199 (1989).

\bibitem{Jain.Book.2007} J. K. Jain, \textit{Composite fermions}, (Cambridge University Press, Cambridge, England, 2007).

\bibitem{Willett.PRL.1987} R. Willett, J. P. Eisenstein, H. L. Störmer, D. C. Tsui, A. C. Gossard, and J. H. English, Observation of an even-denominator quantum number in the fractional quantum Hall effect, Phys. Rev. Lett. {\bf 59}, 1776 (1987).

\bibitem{Lilly.PRL.1999} M. P. Lilly, K. B. Cooper, J. P. Eisenstein, L. N. Pfeiffer, and K. W. West, Evidence for an Anisotropic State of Two-Dimensional Electrons in High Landau Levels, Phys. Rev. Lett. {\bf 82}, 394 (1999).

\bibitem{Du.SSC.1999} R. R. Du, D. C. Tsui, H. L. Stormer, L. N. Pfeiffer, K. W. Baldwin, and K. W. West, Strongly anisotropic transport in higher two-dimensional Landau levels, Solid State Commun. {\bf 109}, 389 (1999).

\bibitem{Eisenstein.PRL.2002} J. P. Eisenstein, K. B. Cooper, L. N. Pfeiffer, and K. W. West, Insulating and Fractional Quantum Hall States in the First Excited Landau Level, Phys. Rev. Lett. {\bf 88}, 076801 (2002).

\bibitem{Kumar.PRL.2010} A. Kumar, G. A. Csáthy, M. J. Manfra, L. N. Pfeiffer and K. W. West, Nonconventional Odd-Denominator Fractional Quantum Hall States in the Second Landau Level, Phys. Rev. Lett. {\bf 105}, 246808 (2010).

\bibitem{Fu.PRL.2020} X. Fu, Q. Shi, M. A. Zudov, G. C. Gardner, J. D. Watson, M. J. Manfra, K. W. Baldwin, L. N. Pfeiffer, and K. W. West, Anomalous Nematic States in High Half-Filled Landau Levels, Phys. Rev. Lett. {\bf 124}, 067601 (2020).

\bibitem{Moore.NPB.1991} G. Moore and N. Read, Nonabelions in the fractional quantum Hall effect, Nucl. Phy. B {\bf 360(2-3)}, 362 (1991).

\bibitem{Nayak.RMP.2008} Chetan Nayak, Steven H. Simon, Ady Stern, Michael Freedman, and Sankar Das Sarma, Non-Abelian anyons and topological quantum computation, Rev. Mod. Phys. {\bf 80}, 1083 (2008).

\bibitem{Banerjee.Nature.2018} M. Banerjee, M. Heiblum, V. Umansky, D. E. Feldman, Y. Oreg, and A. Stern, Observation of half-integer thermal Hall conductance, Nature {\bf 559}, 205 (2018).

\bibitem{Halperin.Book.2020} \textit{Fractional Quantum Hall Effects: New Developments}, edited by B. I. Halperin and J. K. Jain (World Scientific, Singapore, 2020).

\bibitem{Willett.PRX.2023}  R. L. Willett, K. Shtengel, C. Nayak, L. N. Pfeiffer, Y. J. Chung, M. L. Peabody, K. W. Baldwin, and K. W. West, Interference Measurements of Non-Abelian $e/4$ $\&$ Abelian $e/2$ Quasiparticle Braiding, Phys. Rev. X {\bf 13}, 011028 (2023). 

\bibitem{Falson.NatPhys.2015} J. Falson, D. Maryenko, B. Friess, D. Zhang, Y. Kozuka, A. Tsukazaki, J. H. Smet, and M. Kawasaki, Even-denominator fractional quantum Hall physics in ZnO, Nat. Phys. {\bf 11}, 347 (2015).

\bibitem{Hossain.PRL.2018} Md. Shafayat Hossain, Meng K. Ma, Y. J. Chung, L. N. Pfeiffer, K. W. West, K. W. Baldwin, and M. Shayegan, Unconventional Anisotropic Even-Denominator Fractional Quantum Hall State in a System with Mass Anisotropy, Phys. Rev. Lett. {\bf 121}, 256601 (2018).

\bibitem{Ki.NanoLett.2014} D. K. Ki, V. I. Fal'ko, D. A. Abanin, and A. F. Morpurgo, Observation of even denominator fractional quantum Hall effect in suspended bilayer graphene, Nano Lett. {\bf 14}, 2135 (2014).

\bibitem{Zibrov.Nature.2017} A. A. Zibrov, C. Kometter, H. Zhou, E. M. spanton, T. Taniguchi, K. Watanabe, M. P. Zaletel, and A. F. Young, Tunable interacting composite fermion phases in a half-filled bilayer-graphene Landau level, Nature (London) {\bf 549}, 360 (2017).

\bibitem{Li.Science.2017} J. I. A. Li, C. Tan, S. Chen, Y. Zeng, T. Taniguchi, K. Watanabe, J. Hone, and C. R. Dean, Even-denominator fractional quantum Hall states in bilayer graphene, Science {\bf 358}, 648 (2017).

\bibitem{Huang.PRX.2022} Ke Huang, Hailong Fu, Danielle Reifsnyder Hickey, Nasim Alem, Xi Lin, Kenji Watanabe, Takashi Taniguchi, and Jun Zhu, Valley Isospin Controlled Fractional Quantum Hall States in Bilayer Graphene, Phys. Rev. X {\bf 12}, 031019 (2022).

\bibitem{Shi.NatNano.2020} Qianhui Shi, En-Min Shih, Martin V. Gustafsson, Daniel A. Rhodes, Bumho Kim, Kenji Watanabe, Takashi Taniguchi, Zlatko Papi\'c, James Hone and Cory R. Dean, Odd- and even-denominator fractional quantum Hall states in monolayer WSe$_2$, Nat. Nanotechnol. {\bf 15}, 569 (2020).

\bibitem{Koulakov.PRL.1996} A. A. Koulakov, M. M. Fogler, and B. I. Shklovskii, Charge Density Wave in Two-Dimensional Electron Liquid in Weak Magnetic Field, Phys. Rev. Lett. {\bf 76}, 499 (1996).

\bibitem{Fogler.PRB.1996} M. M. Fogler, A. A. Koulakov, and B. I. Shklovskii, Ground state of a two-dimensional electron liquid in a weak magnetic field, Phys. Rev. B {\bf 54}, 1853 (1996).

\bibitem{Moessner.PRB.1996} R. Moessner and J. T. Chalker, Exact results for interacting electrons in high Landau levels, Phys. Rev. B {\bf 54}, 5006 (1996).

\bibitem{Fradkin.PRB.1999} E. Fradkin and S. A. Kivelson, Liquid-Crystal Phases of Quantum Hall Systems, Phys. Rev. B {\bf 59}, 8065 (1999).

\bibitem{Fradkin.PRL.2000} Eduardo Fradkin, Steven A. Kivelson, Efstratios Manousakis, and Kwangsik Nho, Nematic Phase of the Two-Dimensional Electron Gas in a Magnetic Field, Phys. Rev. Lett. {\bf 84}, 1892 (2000).

\bibitem{Fradkin.ARCMP.2010} E. Fradkin, S. A. Kivelson, M. J. Lawler, J. P. Eisenstein, and A. P. Mackenzie, Nematic Fermi Fluids in Condensed Matter Physics, Annu. Rev. Condens. Matter Phys. {\bf 1}, 153 (2010).

\bibitem{Sammon.PRB.2019} See also M. Sammon, X. Fu, Yi Huang, M. A. Zudov, B. I. Shklovskii, G. C. Gardner, J. D. Watson,
M. J. Manfra, K. W. Baldwin, L. N. Pfeiffer, and K. W. West, Resistivity anisotropy of quantum Hall stripe phases, Phys. Rev. B {\bf 100} 241303(R) (2019).

\bibitem{Oganesyan.PRB.2001} Vadim Oganesyan, Steven A. Kivelson, and Eduardo Fradkin, Quantum theory of a nematic Fermi fluid, Phys. Rev. B {\bf 64}, 195109 (2001).

\bibitem{Lee.PRL.2018} Kyungmin Lee, Junping Shao, Eun-Ah Kim, F. D. M. Haldane, and Edward H. Rezayi, Pomeranchuk Instability of Composite Fermi Liquids, Phys. Rev. Lett. {\bf 121}, 147601 (2018).

\bibitem{Xia.NatPhys.2011} Jing Xia, J. P. Eisenstein, L. N. Pfeiffer, and K. W. West, Evidence for a fractionally quantized Hall state with anisotropic longitudinal transport, Nat. Phys. {\bf 7}, 845 (2011).

\bibitem{Liu.PRB.2013} Yang Liu, S. Hasdemir, M. Shayegan, L. N. Pfeiffer, K. W. West, and K. W. Baldwin, Evidence for a $\nu=5/2$ fractional quantum Hall nematic state in parallel magnetic fields, Phys. Rev. B {\bf 88}, 035307 (2013).

\bibitem{Wen.AP.1995} X. G. Wen, Topological orders and edge excitations in fractional quantum Hall states, Adv. Phys. {\bf 44}, 405 (1995).

\bibitem{Haldane.PRL.2011} F. D. M. Haldane, Geometrical Description of the Fractional Quantum Hall Effect, Phys. Rev. Lett. {\bf 107}, 116801 (2011).

\bibitem{Mulligan.PRB.2011} Michael Mulligan, Chetan Nayak, and Shamit Kachru, Effective field theory of fractional quantized Hall nematics, Phys. Rev. B {\bf 84}, 195124 (2011).

\bibitem{Maciejko.PRB.2013} J. Maciejko, B. Hsu, S. A. Kivelson, YeJe Park, and S. L. Sondhi, Field theory of the quantum Hall nematic Transition, Phys. Rev. B {\bf 88}, 125137 (2013).

\bibitem{You.PRX.2014} Yizhi You, Gil Young Cho, and Eduardo Fradkin, Theory of Nematic Fractional Quantum Hall States, Phys. Rev. X {\bf 4}, 041050 (2014).

\bibitem{Wan.PRB.2016} X. Wan and K. Yang, Striped quantum Hall state in a half-filled Landau level, Phys. Rev. B {\bf 93}, 201303(R) (2016).

\bibitem{Yang.PRL.2017} Bo Yang, Zi-Xiang Hu, Ching Hua Lee, and Z. Papić, Generalized Pseudopotentials for the Anisotropic Fractional Quantum Hall Effect, Phys. Rev. Lett. {\bf 118}, 146403 (2017).

\bibitem{Regnault.PRB.2017} N. Regnault, J. Maciejko, S. A. Kivelson, and S. L. Sondhi, Evidence of a fractional quantum Hall nematic phase in a microscopic model, Phys. Rev. B {\bf 96}, 035150 (2017).

\bibitem{Santos.PRX.2019} Luiz H. Santos, Yuxuan Wang, and Eduardo Fradkin, Pair-Density-Wave Order and Paired Fractional Quantum Hall Fluids, Phys. Rev. X {\bf 9}, 021047 (2019).

\bibitem{Yang.PRR.2020} Bo Yang, Microscopic theory for nematic fractional quantum Hall effect, Phys. Rev. Research {\bf 2}, 033362 (2020).

\bibitem{Ye.PRB.2024} Dan Ye, Chen-Xin Jiang, Zi-Xiang Hu, Fractional quantum Hall nematics at $\nu=7/3$ in a tilted magnetic field, Phys. Rev. B {\bf 110}, 165123 (2024).

\bibitem{Pu.PRL.2024} Songyang Pu, Ajit C. Balram, Joseph Taylor, Eduardo Fradkin, and Zlatko Papić, Microscopic Model for Fractional Quantum Hall Nematics, Phys. Rev. Lett. {\bf 132}, 236503 (2024).

\bibitem{Wang.PRL.2023} Chengyu Wang, A. Gupta, Y. J. Chung, L. N. Pfeiffer, K. W. West, K. W. Baldwin, R. Winkler, and M. Shayegan, Highly Anisotropic Even-Denominator Fractional Quantum Hall State in an Orbitally Coupled Half-Filled Landau Level, Phys. Rev. Lett. {\bf 131}, 056302 (2023).

\bibitem{Samkharadze.NatPhys.2016} N. Samkharadze, K. A. Schreiber, G. C. Gardner, M. J. Manfra, E. Fradkin, and G. A. Cs\'athy, Observation of a transition from a topologically ordered to a spontaneously broken symmetry phase, Nat. Phys. {\bf 12}, 191 (2016).

\bibitem{Chung.Natmater.2021} Yoon Jang Chung, K. A. Villegas Rosales, K. W. Baldwin, P. T. Madathil, K. W. West, M. Shayegan and L. N. Pfeiffer, Ultra-high-quality two-dimensional electron systems, Nat. Mater. {\bf 20}, 632-637 (2021).

\bibitem{Chung.PRM.2022} Yoon Jang Chung, C. Wang, S. K. Singh, A. Gupta, K. W. Baldwin, K. W. West, R. Winkler, M. Shayegan, and L. N. Pfeiffer, Record-quality GaAs two-dimensional hole systems, Phy. Rev. Mater. {\bf 6}, 034005 (2022).

\bibitem{Gupta.PRM.2024} Adbhut Gupta, C. Wang, S. K. Singh, K. W. Baldwin, R. Winkler, M. Shayegan, and L. N. Pfeiffer, Ultraclean two-dimensional hole systems with mobilities exceeding $10^7$ cm$^2$/Vs, Phys. Rev. Mater. {\bf 8}, 014004 (2024).

\bibitem{SM} See Supplemental Material for additional data and discussions.

\bibitem{Goldman.PRL.1988} V. J. Goldman, M. Shayegan, and D. C. Tsui, Evidence for the Fractional Quantum Hall State at $\nu=1/7$, Phys. Rev. Lett. {\bf 61}, 881 (1988).

\bibitem{Manfra.PRL.2007} M. J. Manfra, R. de Picciotto, Z. Jiang, S. H. Simon, L. N. Pfeiffer, K. W. West, and A. M. Sergent, Impact of Spin-Orbit Coupling on Quantum Hall Nematic Phases, Phys. Rev. Lett. {\bf 98}, 206804 (2007). In this work, a nearly isotropic phase at $\nu=9/2$ and anisotropic phases at $\nu=7/2$, 11/2, and 13/2 were reported. No FQHS was osberved at $\nu=5/2$ and 7/2, although a shallow minimum in longitudinal resistance (in the high resistance direction) was seen at 5/2.

\bibitem{Zhu.PRL.2002} J. Zhu, W. Pan, H. L. Stormer, L. N. Pfeiffer, and K. W. West, Density-Induced Interchange of Anisotropy Axes at Half-Filled High Landau Levels, Phys. Rev. Lett. {\bf 88}, 116803 (2002).

\bibitem{Pollanen.PRB.2015} J. Pollanen, K. B. Cooper, S. Brandsen, J. P. Eisenstein, L. N. Pfeiffer, and K. W. West, Heterostructure symmetry and the orientation of the quantum Hall nematic phases, Phys. Rev. B {\bf 92}, 115410 (2015).

\bibitem{Lilly.PRL.1999.tilt} M. P. Lilly, K. B. Cooper, J. P. Eisenstein, L. N. Pfeiffer, and K. W. West, Anisotropic States of Two-Dimensional Electron Systems in High Landau Levels: Effect of an In-Plane Magnetic Field, Phys. Rev. Lett. {\bf 83}, 824 (1999).


\bibitem{Pan.PRL.1999.tilt} W. Pan, R. R. Du, H. L. Stormer, D. C. Tsui, L. N. Pfeiffer, K. W. Baldwin, and K. W. West, Strongly Anisotropic Electronic Transport at Landau Level Filling Factor $\nu=9/2$ and $\nu=5/2$ under a Tilted Magnetic Field, Phys. Rev. Lett. {\bf 83}, 820 (1999).

\bibitem{Cooper.PRB.2002} K. B. Cooper, M. P. Lilly, J. P. Eisenstein, L. N. Pfeiffer, and K. W. West, Onset of anisotropic transport of two-dimensional electrons in high Landau levels: Possible isotropic-to-nematic liquid-crystal phase transition, Phys. Rev. B {\bf 65}, 241313(R) (2002).

\bibitem{Jungwirth.PRB.1999} T. Jungwirth, A. H. MacDonald, L. Smrčka, S. M. Girvin, Field-tilt anisotropy energy in quantum Hall stripe states, Phys. Rev. B {\bf 60}, 15574 (1999).

\bibitem{Winkler.Book.2003} R. Winkler,
\textit{Spin-Orbit Coupling Effects in Two-Dimensional Electron and Hole Systems}, (Springer, Berlin, 2003).

\bibitem{Liu.PRB.2014} Yang Liu, S. Hasdemir, D. Kamburov, A. L. Graninger, M. Shayegan, L. N. Pfeiffer, K. W. West, K. W. Baldwin, and R. Winkler, Even-denominator fractional quantum Hall effect at a Landau level crossing, Phys.Rev. B {\bf 89}, 165313 (2014).

\bibitem{Lupatini.PRL.2020} M. Lupatini, P. Knüppel, S. Faelt, R. Winkler, M. Shayegan, A. Imamoglu, and W. Wegscheider, Spin Reversal of a Quantum Hall Ferromagnet at a Landau Level Crossing, Phys. Rev. Lett. {\bf 125}, 067404 (2020).

\bibitem{Ma.PRL.2022} Meng K. Ma, Chengyu Wang, Y. J. Chung, L. N. Pfeiffer, K. W. West, K. W. Baldwin, R. Winkler, and M. Shayegan, Robust Quantum Hall Ferromagnetism near a Gate-Tuned $\nu=1$ Landau Level Crossing, Phys. Rev. Lett. {\bf 129}, 196801 (2022).

\bibitem{Ma.preprint.2022} K. K. W. Ma, M. R. Peterson, V. W. Scarola, and K.
Yang, Fractional quantum Hall effect at the filling factor $\nu=5/2$, \textit{Encyclopedia of Condensed Matter Physics}, 2nd ed., edited by T. Chakraborty (Academic Press, Oxford, 2024), pp. 324–365

\bibitem{Read.PRB.1999} N. Read and E. Rezayi, Beyond paired quantum Hall states: Parafermions and incompressible states in the first excited Landau level, Phys. Rev. B {\bf 59}, 8084 (1999).

\bibitem{Zhu.PRL.2015} W. Zhu, S. S. Gong, F. D. M. Haldane, and D. N. Sheng, Fractional Quantum Hall States at $\nu=13/5$ and 12/5 and Their Non-Abelian Nature, Phys. Rev. Lett. {\bf 115}, 126805 (2015).

\bibitem{Pakrouski.PRB.2016} Kiryl Pakrouski, Matthias Troyer, Yang-Le Wu, Sankar Das Sarma, and Michael R. Peterson, Enigmatic 12/5 fractional quantum Hall effect, Phys. Rev. B {\bf 94}, 075108 (2016).

\bibitem{Tsui.Nature.2024} Yen-Chen Tsui, Minhao He, Yuwen Hu, Ethan Lake, Taige Wang, Kenji Watanabe, Takashi Taniguchi, Michael P. Zaletel, and Ali Yazdani, Direct observation of a magnetic-field-induced Wigner crystal, Nature {\bf 628}, 287 (2024).

\bibitem{Hu.NatPhys.2025} Yuwen Hu, Yen-Chen Tsui, Minhao He, Umut Kamber, Taige Wang, Amir S. Mohammadi, Kenji Watanabe, Takashi Taniguchi, Zlatko Papić, Michael P. Zaletel, and Ali Yazdani, High-resolution tunnelling spectroscopy of fractional quantum Hall states, Nat. Phys. {\bf 21}, 716 (2025).

\end{thebibliography}
\end{document}